\documentclass[english]{IEEEtran}
\usepackage[T1]{fontenc}
\usepackage[latin9]{inputenc}
\usepackage{xcolor}
\usepackage{pdfcolmk}
\usepackage{verbatim}
\usepackage{calc}
\usepackage{amsthm}
\usepackage{amsmath}
\usepackage{amssymb}
\usepackage{graphicx}
\usepackage{esint}
\PassOptionsToPackage{normalem}{ulem}
\usepackage{ulem}

\makeatletter

\providecommand{\tabularnewline}{\\}
\providecolor{lyxadded}{rgb}{0,0,1}
\providecolor{lyxdeleted}{rgb}{1,0,0}

\theoremstyle{plain}
\newtheorem{thm}{\protect\theoremname}
\theoremstyle{definition}
\newtheorem{defn}[thm]{\protect\definitionname}
\theoremstyle{remark}
\newtheorem{rem}[thm]{\protect\remarkname}

\usepackage{cite}
\usepackage{times}

\newtheorem{assumption}{Assumption}
\newtheorem{challenge}{Challenge}

\@ifundefined{showcaptionsetup}{}{%
 \PassOptionsToPackage{caption=false}{subfig}}
\usepackage{subfig}
\makeatother

\usepackage{babel}
\providecommand{\definitionname}{Definition}
\providecommand{\remarkname}{Remark}
\providecommand{\theoremname}{Theorem}

\begin{document}

\title{Two-Stage Subspace Constrained Precoding in Massive MIMO Cellular
Systems}

\author{{\normalsize{}An Liu, }\textit{\normalsize{}Member IEEE}{\normalsize{},
and Vincent Lau,}\textit{\normalsize{} Fellow IEEE}{\normalsize{},\\Department
of Electronic and Computer Engineering, Hong Kong University of Science
and Technology}}
\maketitle
\begin{abstract}
We propose a subspace constrained precoding scheme that exploits the
spatial channel correlation structure in massive MIMO cellular systems
to fully unleash the tremendous gain provided by massive antenna array
with reduced channel state information (CSI) signaling overhead. The
MIMO precoder at each base station (BS) is partitioned into an \textit{inner
precoder} and a \textit{Transmit (Tx) subspace control} matrix. The
inner precoder is adaptive to the local CSI at each BS for spatial
multiplexing gain. The Tx subspace control is adaptive to the channel
statistics for inter-cell interference mitigation and Quality of Service
(QoS) optimization. Specifically, the Tx subspace control is formulated
as a QoS optimization problem which involves an SINR chance constraint
where the probability of each user\textquoteright s SINR not satisfying
a service requirement must not exceed a given outage probability.
Such chance constraint cannot be handled by the existing methods due
to the two-stage precoding structure. To tackle this, we propose a
\textit{bi-convex approximation approach}, which consists of three
key ingredients: random matrix theory, chance constrained optimization
and semidefinite relaxation. Then we propose an efficient algorithm
to find the optimal solution of the resulting bi-convex approximation
problem. Simulations show that the proposed design has significant
gain over various baselines.\end{abstract}

\section{Introduction}

Massive MIMO has been regarded as one of the key technologies in future
wireless networks. In massive MIMO cellular systems, each BS is equipped
with $M\gg1$ antennas. This large spatial degree of freedom (DoF)
of massive MIMO systems can be exploited to significantly increase
the spectrum and energy efficiency \cite{Marzetta_SPM12_LargeMIMO}.
Specifically, there are two important roles for the spatial DoF introduced
by the massive MIMO, namely the\textit{ intra-cell spatial multiplexing}
and the \textit{inter-cell interference mitigation}. However, there
are several practical issues towards achieving the huge performance
gain predicted by the theoretical analysis in massive MIMO systems.
First, to realize the \textit{spatial multiplexing gains} (combating
intra-cell interference) within each BS using conventional MU-MIMO
precoding \cite{Peel_TOC05_RCI,Gershman_SPM2010_SDRBF}, real-time
CSIT (i.e, channel state information at the BS) is required. In most
of the existing works, Time-Division Duplex (TDD) is assumed and channel
reciprocity can be exploited to obtain CSIT via uplink pilot training.
However, in this paper, we are focusing on a more challenging but
important Frequency-Division Duplex (FDD) mode of Massive MIMO. In
this case, channel reciprocity cannot work and the channel estimation
has to be obtained via downlink channel estimation and channel feedback,
which is practically infeasible because the number of independent
pilot symbols available for channel estimation is fundamentally limited
by the channel coherence time and it may become much smaller than
$M$ as $M$ grows large. Second, to realize the \textit{inter-cell
interference mitigation gains} in massive MIMO using coordinated beamforming
\cite{Foschini_POC06_CordMIMO} or cooperative MIMO \cite{somekh2009cooperative},
\textit{real-time global CSIT} knowledge is required for cooperative
MIMO and at least partial global CSIT (i.e., each BS needs to know
the channels from this BS to all users) is required for coordinated
beamforming. However, the acquisition of (partial) global CSIT is
a very challenging problem in practical massive MIMO systems because
both the downlink pilot training and the uplink CSI feedback overhead
will become unacceptable as the number of antennas $M$ grows large.
It is even more difficult to obtain real-time (partial) global CSIT
due to the backhaul signaling latency.

In this paper, we propose a two stage \textit{subspace constrained
precoding} for massive MIMO cellular systems. In practice, due to
local scattering effects \cite{Abdi_JSAC02_localscatering,Zhang_TWC07_localscattering},
the channel vectors of users are usually concentrated in a subspace
with a much smaller dimension than $M$ when $M$ is large. The proposed
two-stage precoding takes advantage of this limited scattering property
to achieve \textit{intra-cell spatial multiplexing} and \textit{inter-cell
interference mitigation} without all the aforementioned practical
issues. Specifically, the MIMO precoder at each BS is partitioned
into an \textit{inner precoder} and a semi-unitary \textit{Tx subspace
control matrix}. The inner precoder is adaptive to real-time local
CSIT per BS for intra-cell spatial multiplexing gain. The Tx subspace
control matrix is adaptive to long-term channel statistics to achieve
the best tradeoff between direct link diversity gain and cross link
(inter-cell) interference mitigation such that the QoS of the users
is maximized. Such subspace constrained precoding structure simultaneously
resolves the aforementioned practical challenges. For instance, the
issue of insufficient pilot symbols for real-time local CSI estimation
is resolved because the BS only needs to estimate the CSI within the
subspace determined by the Tx subspace control, which is of a much
smaller dimension than $M$. Furthermore, the Tx subspace control
is adaptive to the long-term channel statistics, which is insensitive
to the backhaul latency. As a result, the proposed subspace constrained
precoding fully exploits the large number of antennas to simultaneously
achieve intra-cell spatial multiplexing per BS and inter-cell interference
mitigation without an expensive backhaul signaling requirement. 

\begin{table*}
\begin{centering}
{\small{}}%
\begin{tabular}{|l|l|l|}
\hline 
 & \textbf{\footnotesize{}Proposed two stage precoding} & \textbf{\footnotesize{}Two stage precoding in}{\footnotesize{} \cite{Caire_TIT13_JSDM} }\tabularnewline
\hline 
\textbf{\footnotesize{}System topology} & {\footnotesize{}Multi-cell massive MIMO system} & {\footnotesize{}Single cell massive MIMO system}\tabularnewline
\hline 
\textbf{\footnotesize{}Roles of two-stage precoding} & {\footnotesize{}Intra-cell \& Inter-cell interference mitigation} & {\footnotesize{}Intra-cell interference mitigation}\tabularnewline
\hline 
\textbf{\footnotesize{}Subspace dimension partitioning} & {\footnotesize{}Dynamic} & {\footnotesize{}Static}\tabularnewline
\hline 
\textbf{\footnotesize{}Adaptive to heterogeneous path loss } & {\footnotesize{}Yes} & {\footnotesize{}No}\tabularnewline
\textbf{\footnotesize{}and QoS requirements explicitly} &  & \tabularnewline
\hline 
\textbf{\footnotesize{}Design approach} & {\footnotesize{}Non-trivial design based on optimization} & {\footnotesize{}Simple design based on BD method}\tabularnewline
\hline 
\textbf{\footnotesize{}Performance} & {\footnotesize{}Good } & {\footnotesize{}Moderate}\tabularnewline
\hline 
\end{tabular}
\par\end{centering}{\small \par}

{\small{}\protect\caption{\label{tab:Comp8}{\small{}Summary of the differences compared with
existing (state-of-the-art) two stage precoding scheme in }\cite{Caire_TIT13_JSDM}{\small{}.}\textcolor{blue}{\small{} }}
}
\end{table*}

In \cite{Caire_TIT13_JSDM}, a similar two-stage precoding structure
was proposed for single cell massive MIMO systems, where a simple
pre-beamforming (with a pre-determined dimension) is used to control
intra-cell interference and achieve spatial multiplexing gain based
on block diagonalization (BD). However, the solution in \cite{Caire_TIT13_JSDM}
for single cell systems is fundamentally different from the proposed
solution for multi-cell systems with heterogeneous path loss and heterogeneous
QoS requirements, as summarized in Table \ref{tab:Comp8}. For example,
the BD-based pre-beamforming design in \cite{Caire_TIT13_JSDM} does
not explicitly take into account the heterogeneous path loss and QoS
requirements and thus may be far from optimal as shown in the simulations.
Moreover, the pre-beamforming design in \cite{Caire_TIT13_JSDM} is
based on a heuristic (BD) method and the dimensions of the pre-beamforming
matrices for different user clusters are pre-determined and fixed.
In this paper, the design of Tx subspace control is formulated as
a QoS optimization problem with individual QoS requirements and the
dimensions of Tx subspace control matrices are also included in the
optimization. As shown in the simulations, in multi-cell systems with
heterogeneous path loss and QoS requirements, it is very important
to dynamically allocate the dimensions of the Tx subspace control
matrices over different user clusters. With optimal dynamic subspace
dimension partitioning, the proposed solution can achieve a much better
performance than the two-stage precoding in \cite{Caire_TIT13_JSDM}
with static subspace dimension partition. However, the optimization
based solution in this paper also causes several non-trivial technical
challenges. 
\begin{itemize}
\item \textbf{SINR Chance} \textbf{Constraint under Two-Stage Precoding:}
The QoS optimization problem involves an SINR chance constraint where
the probability of each user\textquoteright s SINR not satisfying
a service requirement must not exceed a given outage probability.
Such SINR chance constraint does not have a closed form expression.
There are various \textit{Bernstein techniques} to derive a safe convex
approximation for the chance constraints of standard forms \cite{Bechar_arxiv11_Bernstein-type,Wang_ICASSP11_Bernstein-type}.
However, due to the two-stage precoding structure, the associated
SINR chance constraint does not belong to any of the standard forms.
\item \textbf{Combinatorial Nature of Dimension Partitioning:} Due to the
combinatorial nature, the optimization of subspace dimension partitioning
is very challenging and brute force solution has high complexity for
large $M$.
\end{itemize}

In this paper, we address the above technical challenges using a novel
\textit{bi-convex approximation approach}. The first challenge is
addressed using the \textit{interference approximation }and \textit{SINR
chance constraint restriction}, where we combine the techniques in
random matrix theory and chance constraint optimization to construct
a deterministic and (asymptotically) conservative approximation for
the SINR chance constraint. The second challenge is addressed using
the \textit{semidefinite relaxation (SDR)}, where we apply the SDR
technique to obtain a relaxed bi-convex problem which does not explicitly
involve the combinatorial optimization w.r.t. the\textbf{ }dimension
partitioning variable. We further prove the tightness of the SDR using
the specific structure of the Tx subspace control problem. Finally,
we propose a low complexity iterative algorithm to solve this bi-convex
problem and find the Tx subspace control solution by combining the
convex optimization method and the bisection search method. Simulations
show that the proposed design achieves significant performance gain
compared with various state-of-the-art baselines under various backhaul
signaling latency.

\textit{Notation}\emph{s}: The superscripts $\left(\cdot\right)^{T}$
and $\left(\cdot\right)^{\dagger}$ denote transpose and Hermitian
respectively. For a set $\mathcal{S}$, $\left|\mathcal{S}\right|$
denotes the cardinality of $\mathcal{S}$. The operator $\textrm{diag}\left(\mathbf{a}\right)$
represents a diagonal matrix whose diagonal elements are the elements
of vector $\mathbf{a}$. For a set of $K$ vectors $\mathbf{a}_{i}\in\mathbb{C}^{M},i=1,...,K$
and a subset $\mathcal{S}\subseteq\left\{ 1,...,K\right\} $, the
notation $\left[\mathbf{a}\right]_{i\in\mathcal{S}}$ denote a $M\times\left|\mathcal{S}\right|$
matrix whose columns are drawn from the vectors in $\left\{ \mathbf{a}_{i},i\in\mathcal{S}\right\} $.
The notation $\mathbb{U}^{M\times N}$ denotes the set of all $M\times N$
semi-unitary matrices. For a matrix $\mathbf{A}$, $\textrm{diag}\left(\mathbf{A}\right)$
represents a diagonal matrix whose diagonal elements are the diagonal
elements of $\mathbf{A}$. $\textrm{span}\left(\mathbf{A}\right)$
represents the subspace spanned by the columns of a matrix $\mathbf{A}$.
$\left\Vert \mathbf{A}\right\Vert $ is the spectral radius of $\mathbf{A}$.

\section{System Model\label{sec:System-Model}}

\subsection{Massive MIMO Cellular System}

\begin{figure}
\begin{centering}
\subfloat[A massive MIMO cellular system with $2$ BSs and $8$ users.]{\centering{}\includegraphics[width=60mm]{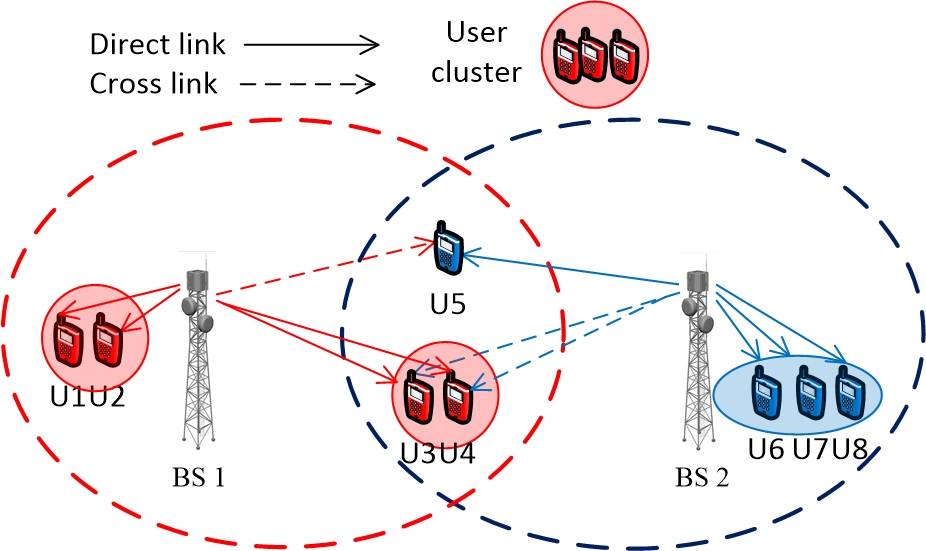}}
\par\end{centering}

\begin{centering}
\subfloat[The corresponding topology graph $\mathcal{G}_{T}\left(\mathbf{\Theta}\right)=\left\{ \mathcal{B},\mathcal{U},\mathcal{E}\right\} $,
where $\mathcal{B}=\left\{ 1,2\right\} $, $\mathcal{U}=\left\{ 1,2,3,4,5,6,7,8\right\} $.]{\centering{}\includegraphics[width=60mm]{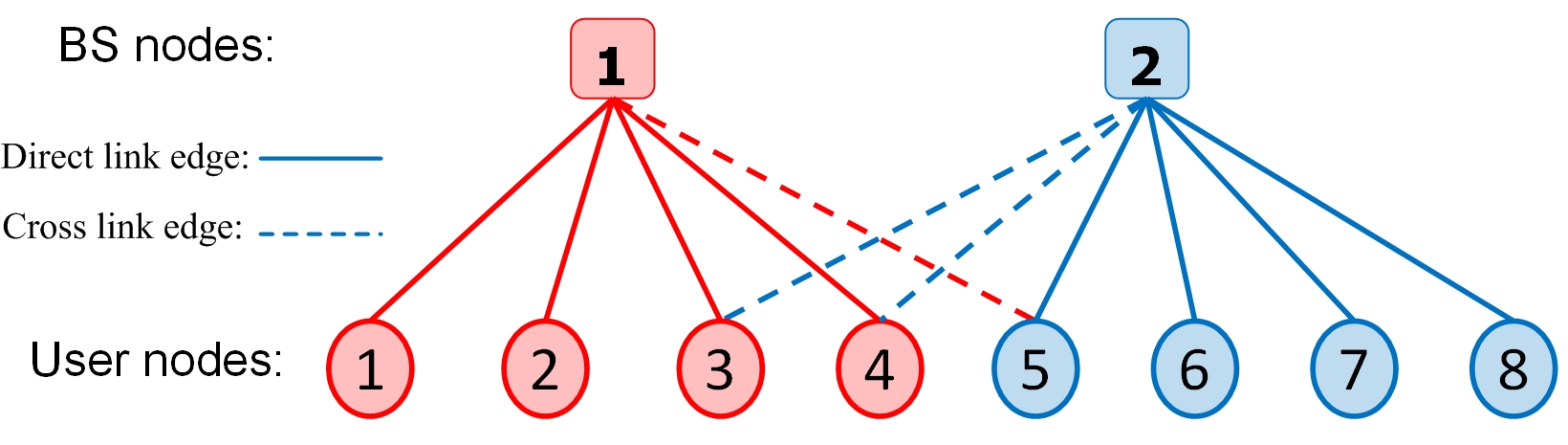}}
\par\end{centering}

\protect\caption{\label{fig:sysmodel}{\small{}An example of a massive MIMO cellular
system and the corresponding topology graph.}}
\end{figure}

Consider the downlink of a massive MIMO cellular system with $L$
BSs and $K$ single-antenna users as illustrated in Fig. \ref{fig:sysmodel}
for $L=2$ and $K=8$. Each BS has $M$ antennas with $M$ much larger
than the number of the associated users. The massive MIMO cellular
system can be represented by a topology graph as define below.

\begin{defn}
[System Topology Graph] Define the \textit{topology graph} of the
massive MIMO cellular system as a bipartite graph $\mathcal{G}_{T}=\left\{ \mathcal{B},\mathcal{U},\mathcal{E}\right\} $,
where $\mathcal{B}$ denotes the set of all BS nodes, $\mathcal{U}$
denotes the set of all user nodes, and $\mathcal{E}$ is the set of
all edges between the BSs and users. An edge $\left(k,l\right)\in\mathcal{E}$
between BS node $l\in\mathcal{B}$ and user node $k\in\mathcal{U}$
represents a wireless link between them. Each edge $\left(k,l\right)\in\mathcal{E}$
is associated with a CSI label $\mathbf{h}_{k,l}\in\mathbb{C}^{M}$,
which represents the channel vector between BS $l$ and user $k$.
For each user node $k$, let $b_{k}$ denote the serving BS.\hfill \IEEEQED
\end{defn}

An example of a massive MIMO cellular system and the corresponding
topology graph is illustrated in Fig. \ref{fig:sysmodel}.

\subsection{Massive MIMO Channel Model }

The channel from BS $l$ to user $k$ is modeled as $\mathbf{h}_{k,l}=\mathbf{\Theta}_{k,l}^{1/2}\mathbf{h}_{k,l}^{w},\:\forall k,l$,
where $\mathbf{h}_{k,l}^{w}\in\mathbb{C}^{M}$ has i.i.d. complex
entries of zero mean and unit variance; and $\mathbf{\Theta}_{k,l}\in\mathbb{C}^{M\times M}$
is a positive semi-definite spatial correlation matrix between BS
$l$ and user $k$. Assume block fading channel where $\mathbf{h}_{k,l}^{w}$
is fixed for a time slot but changes over time slots. As such, the
CSI is divided into instantaneous CSI $\mathbf{H}=\left\{ \mathbf{h}_{k,l}\right\} $
and global statistical information $\mathbf{\Theta}\triangleq\left\{ \mathbf{\Theta}_{k,l}\right\} $
(spatial correlation matrices). We make the following assumption on
the spatial correlation matrices.

\begin{assumption}[Spatial Correlation Model]\label{asm:Chmodel}The
spatial channel correlation matrices satisfy the following two assumptions.
\begin{enumerate}
\item The number of dominant eigenvalues (i.e., the eigenvalues that are
comparable to the maximum eigenvalue) of $\mathbf{\Theta}_{k,l,},\forall k,l$
is relatively small compared to the number of antennas $M$.
\item The users can be grouped into $N$ user clusters $\mathcal{U}_{n},n=1,...,N$
such that all users in $\mathcal{U}_{n}$ are associated with the
same BS denoted by $\overline{l}_{n}$ and $\mathbf{\Theta}_{k,\overline{l}_{n}}=L_{k,\overline{l}_{n}}\mathbf{\Theta}_{n}^{\circ},\forall k\in\mathcal{U}_{n}$,
where $L_{k,\overline{l}_{n}}>0$ is the path loss between BS $\overline{l}_{n}$
and user $k$, and $\textrm{Tr}\left(\mathbf{\Theta}_{n}^{\circ}\right)=M$
is a normalized spatial correlation matrix. 
\end{enumerate}
\end{assumption}

Many experimental measurements as reported in \cite{Kermoal_JSAC02_MIMOdata,Chizhik_JSAC13_MIMOdata}
show that practical MIMO channels indeed have spatial channel correlation
due to limited scattering or LOS transmission. The standard channel
models such as IMT-advanced channel model \cite{ITU_2009_channel}
also predicts that MIMO channel can have high spatial correlation
especially when the angular spread is small. Furthermore, in a massive
MIMO system with a large number of antennas assembled within a limited
space at the BS, the channels are more likely to be highly correlated
due to insufficient spacing among the antennas \cite{Marzetta_SPM12_LargeMIMO,Wang_IJAP14_massiveMIMO}.
As a result, the first assumption has been made in many existing works
on massive MIMO systems, see, e.g., \cite{Wagner_TIT12s_LargeMIMO,Debbah_JSAC12_MCellRMT,Caire_TIT13_JSDM,Gesbert_JSAC13_MassiveMIMOCE,Caire_Arxiv13_JSDMgroup,Wang_IJAP14_massiveMIMO}.
In practice, we can also cluster users into groups using the user
clustering algorithm in \cite{Caire_Arxiv13_JSDMgroup} such that
the second assumption can be closely achieved \cite{Caire_TIT13_JSDM}.
Note that the above channel model is more general than that considered
in \cite{Caire_TIT13_JSDM}. For example, the $n$-th cluster $\mathcal{U}_{n}$
is allowed to contain a single user and the users in the same cluster
can have different path loss. 

At each time slot, linear precoding is employed at BS $\overline{l}_{n}$
to support simultaneous downlink transmissions to the associated users
in $\mathcal{U}_{n}$. Let $\overline{n}_{k}$ denote the index of
the cluster that contains user $k$ and let $\mathcal{B}_{k}=\left\{ n:\: n\neq\overline{n}_{k},\left(k,\overline{l}_{n}\right)\in\mathcal{E}\right\} $
denote the set of user clusters that interfere with user $k$. Then
the received signal for user $k$ can be expressed as:
\begin{eqnarray}
y_{k} & = & \mathbf{h}_{k,b_{k}}^{\dagger}\sqrt{P_{k}}\mathbf{v}_{k}s_{k}+\underset{\textrm{intra-cluster\:\ interference}}{\underbrace{\mathbf{h}_{k,b_{k}}^{\dagger}\sum_{k^{'}\in\mathcal{U}_{\overline{n}_{k}}\backslash\{k\}}\sqrt{P_{k^{'}}}\mathbf{v}_{k^{'}}s_{k^{'}}}}\nonumber \\
 &  & +\underset{\textrm{inter-cluster\:\ interference}}{\underbrace{\sum_{n\in\mathcal{B}_{k}}\mathbf{h}_{k,\overline{l}_{n}}^{\dagger}\mathbf{V}_{n}\mathbf{P}_{n}^{1/2}\mathbf{s}_{n}}}+z_{k},\label{eq:Rxsignal}
\end{eqnarray}
where $s_{k}\sim\mathcal{CN}\left(0,1\right)$ is the data symbol
for user $k$; $\mathbf{v}_{k}\in\mathbb{C}^{M}$ with $\left\Vert \mathbf{v}_{k}\right\Vert =1$
is the precoding vector for user $k$; $P_{k}$ is the power allocated
to user $k$; $\mathbf{s}_{n}=\left[s_{l}\right]_{l\in\mathcal{U}_{n}}\in\mathbb{C}^{\left|\mathcal{U}_{n}\right|}$
is the data symbol vector at BS $n$; $\mathbf{V}_{n}=\left[\mathbf{v}_{l}\right]_{l\in\mathcal{U}_{n}}\in\mathbb{C}^{M\times\left|\mathcal{U}_{n}\right|}$
is the precoding matrix at BS $n$; $\mathbf{P}_{n}=\textrm{diag}\left(\left[P_{l}\right]_{l\in\mathcal{U}_{n}}\right)\in\mathbb{R}_{+}^{\left|\mathcal{U}_{n}\right|\times\left|\mathcal{U}_{n}\right|}$
is the power allocation matrix for the $n$-th user cluster; and $z_{k}\sim\mathcal{CN}\left(0,1\right)$
is the AWGN noise.

\subsection{Two-stage Subspace Constrained Precoding}

The precoder $\mathbf{V}_{n}$ for the $n$-th cluster at BS $\overline{l}_{n}$
is assumed to have a two-stage structure $\mathbf{V}_{n}=\mathbf{F}_{n}\mathbf{G}_{n}$.
The\textit{ Tx subspace control variable} $\mathbf{F}_{n}\in\mathbb{U}^{M\times S_{n}}$
is used to mitigate the inter-cluster interference (including inter-cell
interference) in massive MIMO systems, where $S_{n}\in\left\{ \left|\mathcal{U}_{n}\right|,...,M\right\} $
is called the \textit{dimension partitioning variable }for the $n$-th
cluster. For convenience, define $\mathbf{F}=\left\{ \mathbf{F}_{1},...,\mathbf{F}_{N}\right\} $
as the set of Tx subspace control variables for all user clusters.
Both $\mathbf{F}$ and $S_{n},\forall n$ are computed at a \textit{central
node} based on the global statistical information $\mathbf{\Theta}$.
The inner precoder $\mathbf{G}_{n}=\left[\mathbf{g}_{k}\right]_{k\in\mathcal{U}_{n}}\in\mathbb{C}^{S_{n}\times\left|\mathcal{U}_{n}\right|}$
is used to realize the intra-cluster spatial multiplexing gain using
simple zero-forcing (ZF) precoding. Specifically, for a user $k\in\mathcal{U}_{n}$,
define $\widetilde{\mathbf{h}}_{k,n}\triangleq\mathbf{F}_{n}^{\dagger}\mathbf{h}_{k,\overline{l}_{n}}\in\mathbb{C}^{S_{n}}$
as its \textit{effective channel}. Then, the ZF inner precoding vector
$\mathbf{g}_{k}$ with $\left\Vert \mathbf{g}_{k}\right\Vert =1$
is obtained by the projection of the effective channel $\widetilde{\mathbf{h}}_{k,n}$
on the orthogonal complement of the subspace $\textrm{span}\left(\left[\widetilde{\mathbf{h}}_{k^{'},n}\right]_{k^{'}\in\mathcal{U}_{n}\backslash\left\{ k\right\} }\right)$.
Hence, $\mathbf{G}_{n}$ is computed locally at BS $\overline{l}_{n}$
based on the local real-time instantaneous CSI $\widetilde{\mathbf{H}}_{n}=\left[\widetilde{\mathbf{h}}_{k,n}\right]_{k\in\mathcal{U}_{n}}^{\dagger}\in\mathbb{C}^{\left|\mathcal{U}_{n}\right|\times S_{n}}$.
Note that in the proposed two stage precoding, only the inner precoder
is based on ZF techniques for analytical tractability. The Tx subspace
control is not based on ZF techniques but based on the QoS optimization
problem $\mathcal{P}$ in Section \ref{sec:Optimization-Formulation-for}. 

The diversity gain of the users in the $n$-th cluster increases with
the dimension partitioning variable $S_{n}$. On the other hand, the
interference leakage to other user clusters also increases with $S_{n}$.
As a result, there is a fundamental tradeoff between direct link diversity
gain and cross link interference leakage. It is important to dynamically
adapt the dimension partitioning variable $S_{n}$ based on the spatial
channel correlation matrices $\mathbf{\Theta}$ to optimize the overall
performance. In general, the simple BD-based subspace precoder $\mathbf{F}_{n}$
may be far from optimal as will shown in the simulations. We shall
formulate the design of the Tx subspace control $\mathbf{F}$ and
\textit{subspace dimension partitioning} $S_{n},\forall n$ formally
in Section \ref{sec:Optimization-Formulation-for}.

\section{Optimization Formulation for Tx Subspace and Dimension Partitioning\label{sec:Optimization-Formulation-for}}

Assume that user $k$ has perfect knowledge of the effective single-input
single-output (SISO) channel $\mathbf{h}_{k,b_{k}}^{\dagger}\mathbf{F}_{\overline{n}_{k}}\mathbf{g}_{k}$
and the interference-plus-noise power. Then for given Tx subspace
control $\mathbf{F}$, subspace dimensions $\left\{ S_{n}\right\} $,
and instantaneous CSI $\mathbf{H}$, the SINR of user $k$ is 
\begin{equation}
\textrm{SINR}_{k}=\frac{P_{k}\left|\mathbf{h}_{k,b_{k}}^{\dagger}\mathbf{F}_{\overline{n}_{k}}\mathbf{g}_{k}\right|^{2}}{\sum_{n\in\mathcal{B}_{k}}\mathbf{h}_{k,\overline{l}_{n}}^{\dagger}\mathbf{F}_{n}\mathbf{G}_{n}\mathbf{P}_{n}\mathbf{G}_{n}^{\dagger}\mathbf{F}_{n}^{\dagger}\mathbf{h}_{k,\overline{l}_{n}}+1}.\label{eq:SINRk}
\end{equation}
where $\mathbf{G}_{n}$ is the ZF inner precoder given by a function
of $\mathbf{F}_{n}$ and $\mathbf{H}$ (or more precisely, the effective
channels $\widetilde{\mathbf{H}}_{n}$). In this paper, we focus on
designing Tx subspace control $\mathbf{F}$ to optimize the QoS of
the users. Specifically, we consider the following QoS optimization
problem:
\begin{eqnarray}
 & \mathcal{P}:\:\:\underset{\mathbf{F},\left\{ S_{n}\right\} ,\gamma\geq\gamma^{\circ}}{\max}\gamma\nonumber \\
\textrm{s.t.} & \Pr\left\{ \textrm{SINR}_{k}\geq w_{k}\gamma\right\} \geq1-\epsilon_{k},\forall k;\label{eq:ProbSINR}\\
 & \mathbf{F}_{n}\in\mathbb{U}^{M\times S_{n}},S_{n}\in\left\{ \left|\mathcal{U}_{n}\right|,...,M\right\} ,\forall n,\label{eq:Fncon}
\end{eqnarray}
where the SINR chance constraint in (\ref{eq:ProbSINR}) ensures that
the probability of $\textrm{SINR}_{k}$ not satisfying a service requirement
$w_{k}\gamma$ is less than a maximum allowable outage probability
$\epsilon_{k}\in\left(0,1\right)$, and $w_{k}>0$ is the QoS weight
for user $k$. The constraint $\gamma\geq\gamma^{\circ}$ is used
to guarantee some minimum QoS for all users. The constraint $S_{n}\geq\left|\mathcal{U}_{n}\right|$
is to ensure that the ZF inner precoding is feasible at each BS. The
QoS constraint in (\ref{eq:ProbSINR}) is especially useful for media
streaming applications where there is a stringent delay requirement
for the media packets requested by the users and thus it is desirable
to maintain a fixed data rate (with high probability) for each user.
The QoS weights $w_{k}$'s and the maximum allowable outage probabilities
$\epsilon_{k}$'s can be used to provide differential QoS for different
users.

Note that the optimization of the Tx subspace control variable $\mathbf{F}_{n}$
in Problem $\mathcal{P}$ includes the optimization of both the dimension
partitioning variable $S_{n}$ (or equivalently, the dimension of
$\mathbf{F}_{n}$) and the value of $\mathbf{F}_{n}$. Both $\mathbf{F}_{n}$'s
and $S_{n}$'s are adaptive to the spatial channel correlation matrices
$\mathbf{\Theta}$, i.e., they are fixed and independent of the instantaneous
CSI $\mathbf{H}$ for given $\mathbf{\Theta}$.
\begin{rem}
[Admission Control]\label{Adimission-Control-and}Note that Problem
$\mathcal{P}$ may not always be feasible, i.e., the minimum QoS requirement
$w_{k}\gamma^{\circ}$ cannot be guaranteed for all users even with
infinite transmit power. In practice, an admission control can be
used to ensure that there is enough resource in the system to guarantee
the minimum QoS for all users. The admission control is a challenging
problem and a systematic design of admission control is out of the
scope of this paper. However, we can use a simple admission control
based on the solution of Problem $\mathcal{P}$. For example, whenever
a new user arrives, the admission control solves Problem $\mathcal{P}$
with all users (including the newly arrival user). If Problem $\mathcal{P}$
is feasible, this user is admissible. Otherwise, this user is not
allowed to access the system. As a result, we can focus on the case
when Problem $\mathcal{P}$ is feasible.
\end{rem}

Problem $\mathcal{P}$ is a very challenging problem. First, the probability
functions in the SINR chance constraint (\ref{eq:ProbSINR}) do not
have closed form expressions. Thus, one may only resort to approximation
methods. Due to the two-stage precoding structure, the inner ZF precoder
$\mathbf{G}_{n}$ is also a function of the Tx subspace control variable
$\mathbf{F}_{n}$. As a result, the existing approximation methods
for the ``standard form'' chance constraints in \cite{Bechar_arxiv11_Bernstein-type,Wang_ICASSP11_Bernstein-type}
cannot be applied here. Second, the dimension partitioning variable
$S_{n}$ also needs to be optimized, which is a difficult combinatorial
problem. Third, the domain $\mathbb{U}^{M\times S_{n}}$ of $\mathbf{F}_{n}$
is non-convex. To tackle the above challenges, we propose a novel
\textit{bi-convex approximation approach}, which consists of three
key ingredients: random matrix theory, chance constrained optimization
and semidefinite relaxation (SDR). Then we propose an efficient algorithm
to find the optimal solution of the resulting bi-convex approximation
problem.

\section{The Bi-convex Approximation Approach}

Since it is difficult to solve Problem $\mathcal{P}$ directly, we
propose to find a sub-optimal solution by solving a bi-convex approximation
of $\mathcal{P}$. The proposed bi-convex approximation contains three
steps, where each step solves one key technical challenge associated
with Problem $\mathcal{P}$.

\subsection{Interference Approximation Step}

To handle the SINR chance constraint in (\ref{eq:ProbSINR}), we need
to find a simple characterization of the distribution of $\textrm{SINR}_{k}$,
which is very difficult because the interference term $I_{k}\triangleq\sum_{n\in\mathcal{B}_{k}}\mathbf{h}_{k,\overline{l}_{n}}^{\dagger}\mathbf{F}_{n}\mathbf{G}_{n}\mathbf{P}_{n}\mathbf{G}_{n}^{\dagger}\mathbf{F}_{n}^{\dagger}\mathbf{h}_{k,\overline{l}_{n}}$
in $\textrm{SINR}_{k}$ depends on the instantaneous channel vectors
$\mathbf{h}_{k,\overline{l}_{n}},n\in\mathcal{B}_{k}$ of all the
cross links of user $k$. To simplify the characterization of $\textrm{SINR}_{k}$,
the interference approximation step is used to address the following
challenge.

\begin{center}
\fbox{\begin{minipage}[t]{0.96\columnwidth}%
\begin{challenge}[Deterministic Approximation of $I_{k}$]Find
a deterministic approximation $\hat{I}_{k}$ of $I_{k}$ such that
$\hat{I}_{k}$ does not depend on the instantaneous channel vectors
$\mathbf{h}_{k,\overline{l}_{n}},n\in\mathcal{B}_{k}$ of the cross
links of user $k$.\end{challenge}\vspace{-5bp}
\end{minipage}}
\par\end{center}

We resort to the random matrix theory to solve the above challenge.
Specifically, we apply the random matrix theory to derive an asymptotic
(and deterministic) upper bound $\hat{I}_{k}$ of $I_{k}$ as $M\rightarrow\infty$
and $\left|\mathcal{U}_{n}\right|\rightarrow\infty,\forall n$. We
then use $\hat{I}_{k}$ as an approximation of $I_{k}$ for finite
but large $M$ and $\left|\mathcal{U}_{n}\right|$'s. Throughout the
paper, the notation $M\rightarrow\infty$ refers to $M\rightarrow\infty$
and $\left|\mathcal{U}_{n}\right|\rightarrow\infty,\forall n$ such
that $0<\underset{M\rightarrow\infty}{\liminf}\left|\mathcal{U}_{n}\right|/M\leq\underset{M\rightarrow\infty}{\limsup}\left|\mathcal{U}_{n}\right|/M<\infty$.
The following assumptions are required in order to derive the asymptotic
upper bound of $I_{k}$.

\begin{assumption}[Technical Assumptions for Interference Approximation]\label{asm:LSFM}All
spatial correlation matrices $\mathbf{\Theta}_{k,l},\forall k,l$
have uniformly bounded spectral norm on $M$, i.e., 
\begin{eqnarray}
\underset{M\rightarrow\infty}{\limsup}\underset{1\leq k\leq K}{\textrm{sup}}\left\Vert \mathbf{\Theta}_{k,l}\right\Vert  & < & \infty,\:\forall l.\label{eq:LFbound}
\end{eqnarray}

\end{assumption}

Assumption \ref{asm:LSFM} is satisfied by many MIMO channel model
and it is a standard assumption in the literature, see e.g., \cite{Wagner_TIT12s_LargeMIMO,Debbah_JSAC12_MCellRMT}.

Under Assumption \ref{asm:LSFM}, we have the following theorem.
\begin{thm}
[Asymptotic Interference Upper Bound]\label{thm:Iapprox}Under Assumption
\ref{asm:LSFM} and a feasible Tx subspace control $\mathbf{F}$ (i.e.,
$\mathbf{F}$ satisfies the constraints in (\ref{eq:ProbSINR}) and
(\ref{eq:Fncon}) for some $\gamma\geq\gamma^{\circ}$), we have $I_{k}-\hat{I}_{k}\overset{a.s}{\leq}0$
as $M\rightarrow\infty$, where 
\[
\hat{I}_{k}\triangleq\sum_{n\in\mathcal{B}_{k}}\overline{P}_{n}\textrm{Tr}\left(\mathbf{F}_{n}^{\dagger}\mathbf{\Theta}_{k,\overline{l}_{n}}\mathbf{F}_{n}\right).
\]
and $\overline{P}_{n}=\sum_{k^{'}\in\mathcal{U}_{n}}P_{k^{'}}$ is
the sum transmit power of the users in the $n$-th cluster.
\end{thm}

Please refer to Appendix \ref{sub:Proof-of-TheoremIapprox} for the
proof. 

Replacing the interference term $I_{k}$ in $\textrm{SINR}_{k}$ using
the asymptotic interference upper bound $\hat{I}_{k}$ in Theorem
\ref{thm:Iapprox}, we obtain an asymptotically safe approximation
of the SINR chance constraint in (\ref{eq:ProbSINR})%
\footnote{Asymptotically safe approximation means that if (\ref{eq:CCapprox})
is satisfied, then the SINR chance constraint in (\ref{eq:ProbSINR})
is satisfied as $M\rightarrow\infty$.%
}:
\begin{equation}
\Pr\left\{ \left|\mathbf{h}_{k,b_{k}}^{\dagger}\mathbf{F}_{\overline{n}_{k}}\mathbf{g}_{k}\right|^{2}\geq\frac{w_{k}\gamma}{P_{k}}\left(\hat{I}_{k}+1\right)\right\} \geq1-\epsilon_{k}.\label{eq:CCapprox}
\end{equation}

\subsection{SINR Chance Constraint Restriction Step}

The asymptotically safe approximate SINR chance constraint in (\ref{eq:CCapprox})
remains intractable although it appears to be relatively easier to
handle than the original counterparts in (\ref{eq:ProbSINR}). The
restriction step aims to find a quadratic conservative approximation
of (\ref{eq:CCapprox}). There are some existing methods that use
the worst-case deterministic constraints to approximate various forms
of chance constraints \cite{Ben-Tal_OperR09_chanceConst,Bechar_arxiv11_Bernstein-type,Wang_ICASSP11_Bernstein-type}.
For example, in the context of transmit beamforming design for MU-MIMO
downlink with imperfect CSI \cite{Wang_ICASSP11_Bernstein-type},
the Bernstein-type inequality was used to construct a convex restriction
of the SINR chance constraint that involves a quadratic function of
the standard complex Gaussian vector. Unlike the conventional transmit
beamforming problem, in our problem, the MIMO precoder is divided
into Tx subspace control $\mathbf{F}_{n}$ and inner precoder $\mathbf{G}_{n}$
which change at different time scales. As a result, the asymptotically
safe approximate SINR chance constraint in (\ref{eq:CCapprox}) does
not belong to any of the standard forms that have been studied in
the existing works. For example, the inner precoding vector $\mathbf{g}_{k}$
is also a function of the random channel vectors and thus $\left|\mathbf{h}_{k,b_{k}}^{\dagger}\mathbf{F}_{\overline{n}_{k}}\mathbf{g}_{k}\right|^{2}$
is not a quadratic function of the standard complex Gaussian vector.
In our solution, the restriction step entails finding a solution to
the following:

\begin{center}
\fbox{\begin{minipage}[t]{0.96\columnwidth}%
\begin{challenge}[Asymptotically Quadratic Restriction of SINR Chance
Constraint (\ref{eq:ProbSINR})]\label{chl:Deterministic-Restriction}Find
a quadratic function $\mathbf{f}_{k}\left(\mathbf{F}\right)$ from
$\mathbf{F}$ to $\mathbb{R}^{2}$ such that if $\mathbf{f}_{k}\left(\mathbf{F}\right)\geq\mathbf{0}$,
then the SINR chance constraint in (\ref{eq:ProbSINR}) is asymptotically
satisfied as $M\rightarrow\infty$\end{challenge}\vspace{-5bp}
\end{minipage}}
\par\end{center}

Our strategy is as follows. First, we show that conditioned on $\widetilde{\mathbf{H}}_{-k}\triangleq\left[\widetilde{\mathbf{h}}_{k^{'},\overline{n}_{k}}\right]_{k^{'}\in\mathcal{U}_{\overline{n}_{k}}\backslash\left\{ k\right\} }$,
the effective channel gain $\left|\mathbf{h}_{k,b_{k}}^{\dagger}\mathbf{F}_{\overline{n}_{k}}\mathbf{g}_{k}\right|^{2}$
of user $k$ can be expressed as a quadratic function of a standard
complex Gaussian vector. Then we apply the standard method in \cite{Bechar_arxiv11_Bernstein-type,Wang_ICASSP11_Bernstein-type}
to construct a quadratic restriction of the chance constraint in (\ref{eq:CCapprox})
conditioned on a given $\widetilde{\mathbf{H}}_{-k}$. Finally, we
use the quadratic restriction of the conditional chance constraint
of (\ref{eq:CCapprox}) with the ``worst case'' $\widetilde{\mathbf{H}}_{-k}$
as a quadratic restriction of the unconditional chance constraint
in (\ref{eq:CCapprox}). Since (\ref{eq:CCapprox}) is an asymptotically
safe approximation of (\ref{eq:ProbSINR}), the result will then be
an asymptotically quadratic restriction of the SINR chance constraint
in (\ref{eq:ProbSINR}). The final result is summarized in the following
theorem. Please refer to Appendix \ref{sub:Proof-of-TheoremCA} for
the detailed proof.
\begin{thm}
[Solution to Challenge \ref{chl:Deterministic-Restriction}]\label{thm:Conservative-Approximation}The
asymptotically safe approximate SINR chance constraint in (\ref{eq:CCapprox})
is satisfied if%
\begin{eqnarray}
 & \left(1-\frac{\sqrt{2\delta_{k}}}{\sigma_{k}}\right)\textrm{Tr}\left(\overline{\mathbf{\Theta}}_{\overline{n}_{k}}^{\circ}\mathbf{F}_{\overline{n}_{k}}\mathbf{F}_{\overline{n}_{k}}^{\dagger}\right)\nonumber \\
 & \geq\gamma\sum_{n\in\mathcal{B}_{k}}\textrm{Tr}\left(\overline{\mathbf{\Theta}}_{k,n}\mathbf{F}_{n}\mathbf{F}_{n}^{\dagger}\right)+c_{k}\left(\gamma\right),\nonumber \\
 & \textrm{Tr}\left(\overline{\mathbf{\Theta}}_{\overline{n}_{k}}^{\circ}\mathbf{F}_{\overline{n}_{k}}\mathbf{F}_{\overline{n}_{k}}^{\dagger}\right)\geq\sigma_{k}^{2}+\left|\mathcal{U}_{\overline{n}_{k}}\right|-1,\label{eq:CC12}
\end{eqnarray}
where $\delta_{k}=\ln\frac{1}{\epsilon_{k}}$, $\sigma_{k}=\frac{\sqrt{2\delta_{k}}+\sqrt{2\delta_{k}+4}}{2}$,
$\overline{\mathbf{\Theta}}_{\overline{n}_{k}}^{\circ}=\frac{1}{\lambda_{\overline{n}_{k}}}\mathbf{\Theta}_{\overline{n}_{k}}^{\circ}$,
$\lambda_{\overline{n}_{k}}$ is the largest eigenvalue of $\mathbf{\Theta}_{\overline{n}_{k}}^{\circ}$,
$\overline{\mathbf{\Theta}}_{k,n}=\frac{w_{k}\overline{P}_{n}}{\lambda_{\overline{n}_{k}}P_{k}}\mathbf{\Theta}_{k,\overline{l}_{n}}$
and $c_{k}\left(\gamma\right)=\frac{w_{k}\gamma}{\lambda_{\overline{n}_{k}}P_{k}}+\left(1-\frac{\sqrt{2\delta_{k}}}{\sigma_{k}}\right)\left(\left|\mathcal{U}_{\overline{n}_{k}}\right|-1\right)$.
\end{thm}

In (\ref{eq:CC12}), $\textrm{Tr}\left(\overline{\mathbf{\Theta}}_{\overline{n}_{k}}^{\circ}\mathbf{F}_{\overline{n}_{k}}\mathbf{F}_{\overline{n}_{k}}^{\dagger}\right)$
and $\sum_{n\in\mathcal{B}_{k}}\textrm{Tr}\left(\overline{\mathbf{\Theta}}_{k,n}\mathbf{F}_{n}\mathbf{F}_{n}^{\dagger}\right)$
can be interpreted as the signal power of the direct link and interference
leakage from the cross links. Hence, the quadratic restriction of
(\ref{eq:CCapprox}) in (\ref{eq:CC12}) captures the key tradeoff
between direct link signal power and cross link interference leakage. 
\begin{rem}
The asymptotically quadratic restriction in (\ref{eq:CC12}) is also
valid under finite $M$. The asymptotic approach based on random matrix
theory has been widely used in wireless communications especially
when it is difficult to directly analyze the original system and the
validity of such approach has been verified by numerous works. In
this paper, we utilize random matrix theory and obtain (\ref{eq:CC12})
as a safe approximation of the original SINR chance constraint when
$M\rightarrow\infty$. It has been shown in \cite{Verdu_RMT} that
the random matrix theory results are pretty good approximation even
when $M$ is small. Hence, (\ref{eq:CC12}) is a safe approximation
of the original SINR chance constraint even under finite $M$ in massive
MIMO (as illustrated in Fig. \ref{fig:SINR-Prob} in Section \ref{sec:Simulation-Results}). 
\end{rem}

According to Theorem \ref{thm:Iapprox} and Theorem \ref{thm:Conservative-Approximation},
(\ref{eq:CC12}) is a quadratic restriction of the original SINR chance
constraint in (\ref{eq:ProbSINR}) for large $M$. Replacing the SINR
chance constraint (\ref{eq:ProbSINR}) in $\mathcal{P}$ with the
quadratic constraint in (\ref{eq:CC12}), we have the following asymptotically
conservative formulation of Problem $\mathcal{P}$:
\begin{eqnarray}
 & \mathcal{P}_{1}:\:\:\underset{\mathbf{F},\left\{ S_{n}\right\} ,\gamma\geq\gamma^{\circ}}{\max}\gamma\nonumber \\
\textrm{s.t.} & (\ref{eq:CC12})\textrm{ is satisfied for all }k,\nonumber \\
 & \mathbf{F}_{n}\in\mathbb{U}^{M\times S_{n}},S_{n}\in\left\{ \left|\mathcal{U}_{n}\right|,...,M\right\} ,\forall n.\label{eq:SUcon}
\end{eqnarray}

\subsection{Semidefinite Relaxation Step}

By combining the techniques in random matrix theory and chance constraint
optimization, we have constructed an asymptotically quadratic restriction
for the complicated SINR chance constraint in (\ref{eq:ProbSINR}).
However, we still need to solve the challenge associated with the
combinatorial optimization of the dimension partitioning variable
$S_{n}$ and the semi-unitary constraint on $\mathbf{F}_{n}$.

\begin{center}
\fbox{\begin{minipage}[t]{0.96\columnwidth}%
\begin{challenge}[Dimension Partitioning and Semi-unitary Constraint]\label{chl:CombUni}Find
a bi-convex problem which is equivalent to Problem $\mathcal{P}_{1}$
and does not involve the combinatorial optimization w.r.t. the dimension
partitioning variable $S_{n}$.\end{challenge}\vspace{-5bp}
\end{minipage}}
\par\end{center}

In the following, we apply SDR to solve the above challenge. It is
easy to see that Problem $\mathcal{P}_{1}$ is equivalent to the following
problem
\begin{eqnarray}
 & \mathcal{P}_{2}:\:\:\underset{\mathbf{W},\gamma\geq\gamma^{\circ}}{\max}\gamma\nonumber \\
\textrm{s.t.} & \left(1-\frac{\sqrt{2\delta_{k}}}{\sigma_{k}}\right)\textrm{Tr}\left(\overline{\mathbf{\Theta}}_{\overline{n}_{k}}^{\circ}\mathbf{W}_{\overline{n}_{k}}\right)\nonumber \\
 & \geq\gamma\sum_{n\in\mathcal{B}_{k}}\textrm{Tr}\left(\overline{\mathbf{\Theta}}_{k,n}\mathbf{W}_{n}\right)+c_{k}\left(\gamma\right),\forall k;\label{eq:CC1-1}\\
 & \mathbf{W}_{n}=\mathbf{W}_{n}^{2},\textrm{Tr}\left(\overline{\mathbf{\Theta}}_{n}^{\circ}\mathbf{W}_{n}\right)\geq\eta_{n},\mathbf{W}_{n}\succeq0,\forall n,\label{eq:SUcon-1}
\end{eqnarray}
where $\mathbf{W}=\left\{ \mathbf{W}_{1},...,\mathbf{W}_{N}\right\} $
with $\mathbf{W}_{n}=\mathbf{F}_{n}\mathbf{F}_{n}^{\dagger}\in\mathbb{C}^{M\times M},n=1,...,N$,
and $\eta_{n}=\max_{k\in\mathcal{U}_{n}}\left(\sigma_{k}^{2}+\left|\mathcal{U}_{n}\right|-1\right)$.
The constraint in (\ref{eq:SUcon-1}) is derived from the constraint
(\ref{eq:SUcon}) in $\mathcal{P}_{1}$ and the second constraint
in (\ref{eq:CC12}) using the fact that $\mathbf{W}_{n}$ can be expressed
as $\mathbf{W}_{n}=\mathbf{F}_{n}\mathbf{F}_{n}^{\dagger}$ with $\mathbf{F}_{n}\in\mathbb{U}^{M\times S_{n}}$
if and only if $\mathbf{W}_{n}=\mathbf{W}_{n}^{2},\textrm{Tr}\left(\mathbf{W}_{n}\right)=S_{n},\mathbf{W}_{n}\succeq0$
(Note that $\textrm{Tr}\left(\overline{\mathbf{\Theta}}_{n}^{\circ}\mathbf{W}_{n}\right)\geq\eta_{n}$
implies that $\textrm{Tr}\left(\mathbf{W}_{n}\right)\geq\left|\mathcal{U}_{n}\right|$).
For any feasible solution $\mathbf{W}$ of $\mathcal{P}_{2}$, we
can calculate the corresponding feasible solution $\mathbf{F}=\left\{ \mathbf{F}_{1},...,\mathbf{F}_{N}\right\} $
and $S_{n},\forall n$ of $\mathcal{P}_{1}$ using eigenvalue decomposition
(EVD) $\mathbf{W}_{n}=\mathbf{F}_{n}\mathbf{I}_{S_{n}}\mathbf{F}_{n}^{\dagger},\forall n$
and $S_{n}=\textrm{Tr}\left(\mathbf{W}_{n}\right),\forall n$ respectively.

Problem $\mathcal{P}_{2}$ is still difficult to solve due to the
non-convex constraint $\mathbf{W}_{n}=\mathbf{W}_{n}^{2}$ in (\ref{eq:SUcon-1})
and the bi-linear term $\gamma\sum_{n\in\mathcal{B}_{k}}\textrm{Tr}\left(\overline{\mathbf{\Theta}}_{k,n}\mathbf{W}_{n}\right)$
in (\ref{eq:CC1-1}). To make the problem tractable, we apply SDR
to obtain a convex relaxation of the constraint $\mathbf{W}_{n}=\mathbf{W}_{n}^{2}$
as $\mathbf{W}_{n}-\mathbf{W}_{n}^{2}\succeq\mathbf{0}$, which is
further equivalent to the following convex constraint
\begin{equation}
\left[\begin{array}{cc}
\mathbf{I}_{M} & \mathbf{W}_{n}\\
\mathbf{W}_{n} & \mathbf{W}_{n}
\end{array}\right]\succeq\mathbf{0}.\label{eq:SDRcon}
\end{equation}
Then by replacing the constraint $\mathbf{W}_{n}=\mathbf{W}_{n}^{2}$
with the relaxed constraint in (\ref{eq:SDRcon}) and removing the
constraint $\gamma\geq\gamma^{\circ}$, we obtain a relaxed problem
of $\mathcal{P}_{2}$ as
\begin{eqnarray}
 & \hat{\mathcal{P}}:\:\:\underset{\mathbf{W},\gamma}{\max}\gamma\nonumber \\
\textrm{s.t.} & a_{k}\textrm{Tr}\left(\overline{\mathbf{\Theta}}_{\overline{n}_{k}}^{\circ}\mathbf{W}_{\overline{n}_{k}}\right)\geq\nonumber \\
 & \gamma\sum_{n\in\mathcal{B}_{k}}\textrm{Tr}\left(\overline{\mathbf{\Theta}}_{k,n}\mathbf{W}_{n}\right)+c_{k}\left(\gamma\right),\forall k\label{eq:CCPh}\\
 & \textrm{Tr}\left(\overline{\mathbf{\Theta}}_{n}^{\circ}\mathbf{W}_{n}\right)\geq\eta_{n},\forall n\label{eq:CC1Ph}\\
 & \left[\begin{array}{cc}
\mathbf{I}_{M} & \mathbf{W}_{n}\\
\mathbf{W}_{n} & \mathbf{W}_{n}
\end{array}\right]\succeq\mathbf{0},\mathbf{W}_{n}\succeq0,\forall n\label{eq:Wcon}
\end{eqnarray}
where $a_{k}=\left(1-\frac{\sqrt{2\delta_{k}}}{\sigma_{k}}\right)$.

Problem $\hat{\mathcal{P}}$ is a bi-convex problem, i.e., it is convex
w.r.t. $\mathbf{W}$ ($\gamma$) for fixed $\gamma$ ($\mathbf{W}$),
but is not jointly convex. Note that Problem $\hat{\mathcal{P}}$
does not involve the combinatorial optimization w.r.t. the dimension
partitioning variable $S_{n}$ because the optimal $S_{n}$ is automatically
determined by the rank of the optimal $\mathbf{W}_{n}$. To completely
solve Challenge \ref{chl:CombUni}, we further prove the equivalence
between $\hat{\mathcal{P}}$ and $\mathcal{P}_{1}$ in the following
theorem. 
\begin{thm}
[Equivalence between $\hat{\mathcal{P}}$ and $\mathcal{P}_{1}$]\label{thm:Tightness-of-SDR}The
optimal solution $\mathbf{W}^{\star},\gamma^{\star}$ of $\hat{\mathcal{P}}$
satisfies $\mathbf{W}_{n}^{\star}-\mathbf{W}_{n}^{\star2}=\mathbf{0},\forall n$.
Moreover, if $\mathcal{P}_{1}$ is feasible, then $\gamma^{\star}\geq\gamma^{\circ}$
and thus $\mathbf{W}^{\star},\gamma^{\star}$ is also optimal for
$\mathcal{P}_{1}$. 
\end{thm}

Please refer to Appendix \ref{Proof-of-TheoremSDR} for the proof.

\subsection{Optimal Solution of Problem $\hat{\mathcal{P}}$ }

In this section, we propose an iterative algorithm named \textit{Algorithm
BCA} to solve $\hat{\mathcal{P}}$ by combining the convex optimization
method (for the optimization of $\mathbf{W}$) and the bisection search
method (for the optimization of $\gamma$). We first propose a low
complexity algorithm to solve $\hat{\mathcal{P}}$ with fixed $\gamma$
based on the Lagrange dual method. Then we summarize the overall Algorithm
BCA and establish its global convergence.

\subsubsection{Lagrange dual method for solving Problem $\hat{\mathcal{P}}$ with
fixed $\gamma$\label{sub:Lagrange-dual-method}}

For fixed $\gamma$, problem $\hat{\mathcal{P}}$ is convex and thus
can be solved by the standard convex optimization method \cite{Boyd_04Book_Convex_optimization}.
However, the computation complexity of standard convex optimization
method is still quite high as the number of antennas $M$ becomes
large. Based on the Lagrange dual method, we exploit the specific
structure of the problem to propose a low complexity algorithm which
can significantly reduce the computation complexity over the standard
convex optimization method. 

The Lagrange function of Problem $\hat{\mathcal{P}}$ with fixed $\gamma$
is
\begin{eqnarray*}
L\left(\boldsymbol{\mu},\boldsymbol{\nu},\mathbf{W}\right) & = & \gamma+\sum_{n=1}^{N}\nu_{n}\left(\textrm{Tr}\left(\overline{\mathbf{\Theta}}_{n}^{\circ}\mathbf{W}_{n}\right)-\eta_{n}\right)+\\
 &  & \sum_{k=1}^{K}\mu_{k}\bigg[a_{k}\textrm{Tr}\left(\overline{\mathbf{\Theta}}_{\overline{n}_{k}}^{\circ}\mathbf{W}_{\overline{n}_{k}}\right)-\\
 &  & \gamma\sum_{n\in\mathcal{B}_{k}}\textrm{Tr}\left(\overline{\mathbf{\Theta}}_{k,n}\mathbf{W}_{n}\right)-c_{k,1}-c_{k,2}\gamma\bigg],
\end{eqnarray*}
where $\boldsymbol{\mu}=\left[\mu_{k}\right]_{k=1,...,K}\in\mathbb{R}_{+}^{K}$
is the Lagrange multiplier vector associated with the constraint in
(\ref{eq:CCPh}), $\boldsymbol{\nu}=\left[\nu_{n}\right]_{n=1,...,N}\in\mathbb{R}_{+}^{N}$
is the Lagrange multiplier vector associated with the constraint in
(\ref{eq:CC1Ph}), $c_{k,1}=a_{k}\left(\left|\mathcal{U}_{\overline{n}_{k}}\right|-1\right)$
and $c_{k,2}=\frac{w_{k}}{\lambda_{\overline{n}_{k}}P_{k}}$. Note
that $L\left(\boldsymbol{\mu},\boldsymbol{\nu},\mathbf{W}\right)$
is a partial Lagrangian since (\ref{eq:Wcon}) is kept as a separate
constraint. The dual function of Problem $\hat{\mathcal{P}}$ with
fixed $\gamma$ is
\begin{equation}
J\left(\boldsymbol{\mu},\boldsymbol{\nu}\right)\triangleq\underset{\mathbf{W}}{\textrm{max}}\: L\left(\boldsymbol{\mu},\boldsymbol{\nu},\mathbf{W}\right),\:\textrm{s.t. }(\ref{eq:Wcon})\textrm{ is satisfied}.\label{eq:DualFun}
\end{equation}
The corresponding dual problem is
\begin{equation}
\underset{\boldsymbol{\mu},\boldsymbol{\nu}}{\textrm{min}}\: J\left(\boldsymbol{\mu},\boldsymbol{\nu}\right),\:\textrm{s.t.}\:\boldsymbol{\mu}\geq\mathbf{0},\boldsymbol{\nu}\geq\mathbf{0}.\label{eq:Dualproblem}
\end{equation}

Note that $L\left(\boldsymbol{\mu},\boldsymbol{\nu},\mathbf{W}\right)=\sum_{n=1}^{N}\textrm{Tr}\left(\mathbf{A}_{n}\mathbf{W}_{n}\right)+c^{'}$,
where $\mathbf{A}_{n}=\sum_{k\in\mathcal{U}_{n}}\mu_{k}a_{k}\overline{\mathbf{\Theta}}_{\overline{n}_{k}}^{\circ}-\gamma\sum_{k\in\overline{\mathcal{U}}_{n}}\mu_{k}\overline{\mathbf{\Theta}}_{k,n}+\nu_{n}\overline{\mathbf{\Theta}}_{n}^{\circ}$,
$\overline{\mathcal{U}}_{n}=\left\{ k:\:\overline{n}_{k}\neq n,\left(k,\overline{l}_{n}\right)\in\mathcal{E}\right\} $,
and $c^{'}$ is independent of $\mathbf{W}$. Then the maximization
problem in (\ref{eq:DualFun}) can be decomposed into $N$ independent
problems as 
\begin{equation}
\underset{\mathbf{W}_{n}}{\textrm{max}}\:\textrm{Tr}\left(\mathbf{A}_{n}\mathbf{W}_{n}\right),\:\textrm{s.t. }\mathbf{W}_{n}-\mathbf{W}_{n}^{2}\succeq\mathbf{0},\mathbf{W}_{n}\succeq0,\label{eq:maxLm}
\end{equation}
for $n=1,...,N$. Let $\mathbf{A}_{n}=\mathbf{U}_{n}\mathbf{D}_{n}\mathbf{U}_{n}^{\dagger}$
be the EVD of $\mathbf{A}_{n}$ and let $\mathbf{U}_{n}^{*}$ denote
the eigenvectors corresponding to the positive eigenvalues of $\mathbf{A}_{n}$.
Then for fixed $\boldsymbol{\mu},\boldsymbol{\nu}$, it can be shown
that Problem (\ref{eq:maxLm}) has a closed form solution given by
\begin{equation}
\mathbf{W}_{n}^{*}\left(\boldsymbol{\mu},\boldsymbol{\nu}\right)=\mathbf{U}_{n}^{*}\mathbf{U}_{n}^{*\dagger}.\label{eq:optWn}
\end{equation}
Since Problem $\hat{\mathcal{P}}$ with fixed $\gamma$ is convex,
the optimal solution is given by $\mathbf{W}^{*}\left(\boldsymbol{\mu}^{*},\boldsymbol{\nu}^{*}\right)=\left\{ \mathbf{W}_{n}^{*}\left(\boldsymbol{\mu}^{*},\boldsymbol{\nu}^{*}\right):\forall n\right\} $,
where $\left(\boldsymbol{\mu}^{*},\boldsymbol{\nu}^{*}\right)$ is
the optimal solution of the equivalent dual problem in (\ref{eq:Dualproblem}).

The dual function $J\left(\boldsymbol{\mu},\boldsymbol{\nu}\right)$
is convex and $\nabla J\left(\boldsymbol{\mu},\boldsymbol{\nu}\right)=\left[\partial_{\mu_{1}}J\left(\boldsymbol{\mu},\boldsymbol{\nu}\right),...,\partial_{\mu_{K}}J\left(\boldsymbol{\mu},\boldsymbol{\nu}\right),\partial_{\nu_{1}}J\left(\boldsymbol{\mu},\boldsymbol{\nu}\right),...,\partial_{\nu_{N}}J\left(\boldsymbol{\mu},\boldsymbol{\nu}\right)\right]^{T}$
is a subgradient of $J\left(\boldsymbol{\mu},\boldsymbol{\nu}\right)$
at $\left(\boldsymbol{\mu},\boldsymbol{\nu}\right)$, where 
\begin{eqnarray}
\partial_{\mu_{k}}J\left(\boldsymbol{\mu},\boldsymbol{\nu}\right) & = & a_{k}\textrm{Tr}\left(\overline{\mathbf{\Theta}}_{\overline{n}_{k}}^{\circ}\mathbf{W}_{\overline{n}_{k}}^{*}\left(\boldsymbol{\mu},\boldsymbol{\nu}\right)\right)-c_{k,2}\gamma\nonumber \\
 &  & -\gamma\sum_{n\in\mathcal{B}_{k}}\textrm{Tr}\left(\overline{\mathbf{\Theta}}_{k,n}\mathbf{W}_{n}^{*}\left(\boldsymbol{\mu},\boldsymbol{\nu}\right)\right)-c_{k,1},\forall k\nonumber \\
\partial_{\nu_{n}}J\left(\boldsymbol{\mu},\boldsymbol{\nu}\right) & = & \textrm{Tr}\left(\overline{\mathbf{\Theta}}_{n}^{\circ}\mathbf{W}_{n}^{*}\left(\boldsymbol{\mu},\boldsymbol{\nu}\right)\right)-\eta_{n},\forall n\label{eq:subJ}
\end{eqnarray}
Hence, the standard subgradient based methods such as the subgradient
algorithm in \cite{Boyd_03note_Subgradient} or the Ellipsoid method
in \cite{Boyd_04Book_Convex_optimization} can be used to solve the
optimal solution $\left(\boldsymbol{\mu}^{*},\boldsymbol{\nu}^{*}\right)$
of the dual problem in (\ref{eq:Dualproblem}).
\begin{rem}
For Problem $\hat{\mathcal{P}}$ with fixed $\gamma$, the Lagrange
dual method is a better choice than other standard convex optimization
methods such as interior point method \cite{Boyd_04Book_Convex_optimization}
for two reasons: 1) the number of primal variables in this problem
is much larger than the number of dual variables ($NM^{2}+1$ versus
$K+N$ with $M\gg K$ and $M\gg N$); 2) for fixed Lagrange multipliers
$\boldsymbol{\mu},\boldsymbol{\nu}$, the optimal primal variables
$\mathbf{W}_{n}^{*}\left(\boldsymbol{\mu},\boldsymbol{\nu}\right)$
have closed form solution. Hence, although the convergence speed of
the simple subgradient and Ellipsoid algorithms for the optimization
of dual variables is not as fast as the more complicated algorithms,
the overall computation time (to reach a given accuracy) of the Lagrange
dual method is much smaller than other standard convex optimization
methods.
\end{rem}

\subsubsection{Summary of the Overall Algorithm for solving Problem $\hat{\mathcal{P}}$ }

Algorithm BCA is summarized in Table \ref{tab:AlgBCA} and the global
convergence is established in the following theorem.
\begin{table}
\protect\caption{\label{tab:AlgBCA}Algorithm BCA (for solving Problem $\hat{\mathcal{P}}$)}

\centering{}%
\begin{tabular}{l}
\hline 
\textbf{\small{}Initialization}{\small{}: Choose $\gamma_{\textrm{min}}=\gamma^{\circ}$
and $\gamma_{\textrm{max}}=\overline{\gamma}^{\circ}>\gamma^{\circ}$
such }\tabularnewline
{\small{}that Problem $\hat{\mathcal{P}}$ with fixed $\gamma=\overline{\gamma}^{\circ}$
is infeasible. Let $i=1$.}\tabularnewline
\textbf{\small{}Step 1:}{\small{} Let $\gamma^{(i)}=\frac{\gamma_{\textrm{min}}+\gamma_{\textrm{max}}}{2}$.
Solve Problem $\hat{\mathcal{P}}$ with }\tabularnewline
{\small{}\ \ \ \ \ \ \ fixed $\gamma=\gamma^{(i)}$ using the
Lagrange dual method }\tabularnewline
{\small{}\ \ \ \ \ \ \ in Section \ref{sub:Lagrange-dual-method}.}\tabularnewline
\textbf{\small{}Step 2:}{\small{} If a feasible solution, denoted
by $\mathbf{W}^{(i)}$, is found}\tabularnewline
{\small{}\ \ \ \ \ \ \ in Step 1, let $\gamma_{\textrm{min}}=\gamma^{(i)}$;
otherwise, let $\gamma_{\textrm{max}}=\gamma^{(i)}$. }\tabularnewline
\textbf{\small{}Step 3:}{\small{} If $\left|\gamma_{\textrm{max}}-\gamma_{\textrm{min}}\right|\leq\varepsilon$,
where $\varepsilon>0$ is}\tabularnewline
{\small{}\ \ \ \ \ \ \ the tolerable error, terminate the algorithm. }\tabularnewline
{\small{}\ \ \ \ \ \ \ Otherwise, let $i=i+1$ and return to
Step 1.}\tabularnewline
\hline 
\end{tabular}
\end{table}

\begin{thm}
[Global Convergence of Algorithm BCA]\label{thm:Global-Convergence-ofBCA}For
fixed tolerable error $\varepsilon$, Algorithm BCA terminates after
$I^{\varepsilon}=\left\lceil \textrm{log}_{2}\left(\frac{\left|\overline{\gamma}^{\circ}-\gamma^{\circ}\right|}{\varepsilon}\right)\right\rceil $
iterations. Let $\mathbf{W}^{\varepsilon}=\mathbf{W}^{\left(I^{\varepsilon}\right)},\gamma^{\varepsilon}=\gamma^{(I^{\varepsilon})}$
denote the solution found by Algorithm BCA with tolerable error $\varepsilon$.
Then we have $\left|\gamma^{\varepsilon}-\gamma^{\star}\right|\leq\varepsilon$.
Moreover, as $\varepsilon\rightarrow0$, we have $\mathbf{W}^{\varepsilon}\rightarrow\mathbf{W}^{\star}$
and $\gamma^{\varepsilon}\rightarrow\gamma^{\star}$.
\end{thm}

\textit{Sketch of Proof:} Algorithm BCA ensures that $\gamma^{\star}\in\left[\gamma_{\textrm{min}},\gamma_{\textrm{max}}\right]$
after each iteration. Moreover, after the $i$-th iteration, we have
$\left|\gamma_{\textrm{max}}-\gamma_{\textrm{min}}\right|\leq\frac{\left|\overline{\gamma}^{\circ}-\gamma^{\circ}\right|}{2^{i}}$
and $\gamma^{(i)}\in\left[\gamma_{\textrm{min}},\gamma_{\textrm{max}}\right]$.
Hence, we have $\left|\gamma^{(i)}-\gamma^{\star}\right|\leq\frac{\left|\overline{\gamma}^{\circ}-\gamma^{\circ}\right|}{2^{i}}$,
from which Theorem \ref{thm:Global-Convergence-ofBCA} follows immediately.\hfill \IEEEQED

\subsection{Summary of the Overall Solution\label{sub:Summary-Solu}}

\begin{figure}
\begin{centering}
\includegraphics[width=85mm]{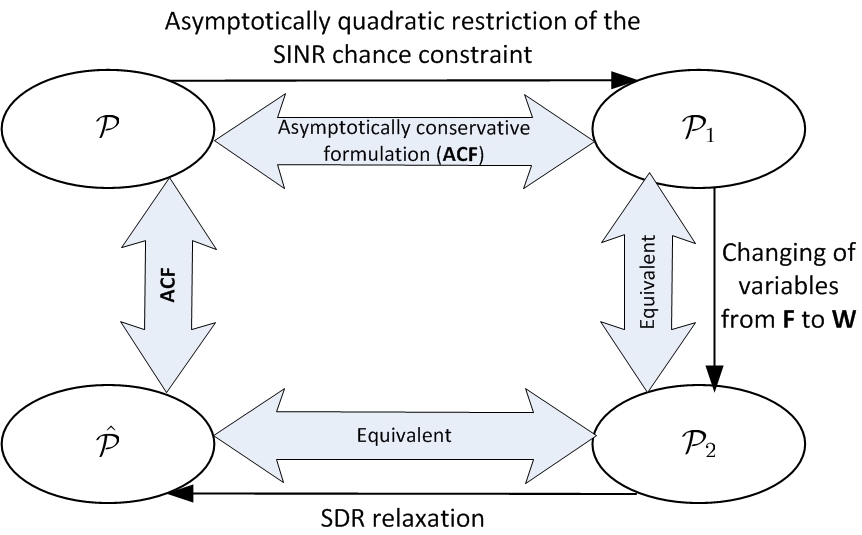}
\par\end{centering}

\protect\caption{\label{fig:prs}{\small{}Illustration of the relationship between
different problems.}}
\end{figure}

The relationship between problems $\mathcal{P}$, $\mathcal{P}_{1}$,
$\mathcal{P}_{2}$ and $\hat{\mathcal{P}}$ is summarized in Fig.
\ref{fig:prs} and is elaborated below.
 
\begin{itemize}
\item \textbf{Relationship between $\mathcal{P}$ and $\mathcal{P}_{1}$:}
Problem $\mathcal{P}_{1}$ is obtained by replacing the original SINR
chance constraint in (\ref{eq:ProbSINR}) with its asymptotically
quadratic restriction in (\ref{eq:CC12}). Hence, Problem $\mathcal{P}_{1}$
is an asymptotically conservative formulation of the Problem $\mathcal{P}$,
i.e., the solution of $\mathcal{P}_{1}$ is a feasible solution to
the original problem $\mathcal{P}$ for sufficiently large $M$. 
\item \textbf{Relationship between $\mathcal{P}$ and $\hat{\mathcal{P}}$:}
By applying the SDR technique, we obtain Problem \textbf{$\hat{\mathcal{P}}$}
as a relaxation of $\mathcal{P}_{1}$. By Theorem \ref{thm:Tightness-of-SDR},
such relaxation is tight and thus \textbf{$\hat{\mathcal{P}}$ }is
equivalent to \textbf{$\mathcal{P}_{1}$. }Hence,\textbf{ $\hat{\mathcal{P}}$
}is also an asymptotically conservative formulation of the Problem
$\mathcal{P}$.
\end{itemize}

Then we summarize the overall solution. First, Algorithm BCA is performed
to calculate the optimal solution $\mathbf{W}^{\star},\gamma^{\star}$
(up to some tolerable error $\varepsilon$) for Problem \textbf{$\hat{\mathcal{P}}$}.
Then we calculate the corresponding Tx subspace control $\mathbf{F}^{\star}=\left\{ \mathbf{F}_{1}^{\star},...,\mathbf{F}_{N}^{\star}\right\} $
using EVD $\mathbf{W}_{n}^{\star}=\mathbf{F}_{n}^{\star}\mathbf{I}_{S_{n}^{\star}}\mathbf{F}_{n}^{\star\dagger},\forall n$,
where $S_{n}^{\star}=\textrm{Tr}\left(\mathbf{W}_{n}^{\star}\right)$.
Finally, we output $\mathbf{F}^{\star},\left\{ S_{n}^{\star}\right\} ,\gamma^{\star}$
as an approximate solution to the original Problem $\mathcal{P}$.
Note that $\mathbf{F}^{\star},\left\{ S_{n}^{\star}\right\} ,\gamma^{\star}$
is usually a conservative solution in the sense that the actual outage
probability of each user $k$ under $\mathbf{F}^{\star},\left\{ S_{n}^{\star}\right\} ,\gamma^{\star}$
is much lower than the maximum allowable outage probability $\epsilon_{k}$. 
\begin{rem}
In this paper, we focus on the optimization of Tx subspace control
$\mathbf{F}$ and subspace dimensions $\left\{ S_{n}\right\} $ for
fixed power control $\left\{ P_{k}\right\} $ (problem $\mathcal{P}$),
which is the key of the two stage precoding design for massive MIMO
systems. The results and algorithm in this paper also provides a basis
for solving the more complicated joint optimization of Tx subspace
control and power control $\left\{ P_{k}\right\} $. For example,
we may design an alternating optimization (AO) algorithm to solve
the joint optimization problem. In each iteration of the AO algorithm,
Algorithm BCA is used as a component to find the optimal Tx subspace
control for fixed power control, and then the optimal power control
is obtained for fixed Tx subspace control.
\end{rem}

\subsection{Methods to Improve the Performance Over the Conservative Solution\label{sub:Methods-to-Improve}}

One common drawback of using the worst-case deterministic constraint
to approximate chance constraint is that, the obtained solution is
usually too conservative \cite{Wang_ICASSP11_Bernstein-type}. In
the context of the probabilistic SINR constrained beamforming problem,
the bisection refinement method has been proposed to mitigate the
conservatism due to the deterministic restriction of the probabilistic
SINR constraint \cite{Wang_ICASSP11_Bernstein-type}. This bisection
refinement method can also be applied in our problem to improve the
performance of the conservative solution obtained in Section \ref{sub:Summary-Solu}.
However, the bisection refinement method in \cite{Wang_ICASSP11_Bernstein-type}
requires solving Problem \textbf{$\hat{\mathcal{P}}$} for multiple
times, which increases the computation complexity. To address this
problem, we propose a simple non-iterative refinement method to determine
the proper value of $\delta_{k}$. First, we calculate the simple
BD-based subspace control $\mathbf{F}^{\textrm{BD}}$ using baseline
3 in Section \ref{sec:Simulation-Results}. Then, we let $\gamma^{\textrm{BD}}=\check{\Phi}^{\textrm{BD}}\left(\epsilon_{k}\right)/w_{k},\forall k$,
where $\Phi^{\textrm{BD}}$ is the empirical CDF of the SINR of user
$k$, denoted by $\textrm{SINR}_{k}^{\textrm{BD}}$, under BD-based
subspace control $\mathbf{F}^{\textrm{BD}}$, and $\check{\Phi}^{\textrm{BD}}$
is the inverse function of $\Phi^{\textrm{BD}}$. Note that $\Pr\left\{ \textrm{SINR}_{k}^{\textrm{BD}}\geq w_{k}\gamma^{\textrm{BD}}\right\} \approx1-\epsilon_{k}$.
Finally, $\delta_{k}$ is chosen as the largest number that satisfies
(\ref{eq:CC12}) with $\mathbf{F}=\mathbf{F}^{\textrm{BD}}$ and $\gamma=\gamma^{\textrm{BD}}$.
Simulations show that the non-iterative refinement method achieves
almost the same performance as the much more complicated bisection
refinement method.

\section{Implementation Considerations}

In this section, we discuss various implementation issues, such as
how to obtain the statistical and real-time CSI, the signaling overhead
and the impact of CSI error. In the following analysis, we assume
that the spatial channel correlation matrices $\mathbf{\Theta}$ change
every $T$ time slots and the rank of each spatial correlation matrix
is $R$.

\subsection{Slow Statistical CSI Acquisition\label{sub:Slow-Statistical-CSI}}

Under typical scenario, the spatial channel correlation matrices change
after a few seconds and the time slot duration is in the order of
millisecond, i.e., $T$ is usually in the order of a few thousands.
For example, in urban areas with a carrier frequency of 2GHz and a
user speed of 3km/h, the spatial channel correlation matrices remain
unchanged for about 22 seconds according to the measurement results
reported in \cite{Ingo_COST02_Svar}. Hence, the statistical CSI acquisition
can be done at a much slower timescale compared to the real-time CSI
acquisition. Specifically, we propose a \textit{rank deficient oracle
approximating shrinkage (OAS) estimator} which is able to obtain an
accurate estimation of the spatial correlation matrix using only $T_{p}=O\left(R\right)\ll M$
independent channel samples.\medskip{}

\textit{Rank deficient OAS estimator:}

\textbf{Input:} $T_{p}$ independent channel samples $\mathbf{h}\left(j\right)\in\mathbb{C}^{M},j=1,...,T_{p}$
of a random vector $\mathbf{h}\in\mathbb{C}^{M}$ with covariance
matrix $\mathbf{\Theta}$.

\textbf{Step 1:} Using Gram\textendash Schmidt process to obtain a
semi unitary matrix $\mathbf{U}\in\mathbb{C}^{M\times R}$ whose columns
form an orthogonal basis for $\mathbf{h}\left(j\right)\in\mathbb{C}^{M},j=1,...,T_{p}$. 

\textbf{Step 2:} Form $T_{p}$ independent channel samples with reduced
dimension $\mathbf{h}^{'}\left(j\right)=\mathbf{U}^{\dagger}\mathbf{h}\left(j\right)\in\mathbb{C}^{R},j=1,...,T_{p}$.
Note that $\mathbf{h}^{'}\left(j\right)$'s can be viewed as realizations
of a random vector $\mathbf{h}^{'}\in\mathbb{C}^{R}$ with covariance
matrix $\mathbf{U}^{\dagger}\mathbf{\Theta}\mathbf{U}$.

\textbf{Step 3:} Use the OAS estimator in \cite{Chen_TSP10_OAS} with
independent samples $\mathbf{h}^{'}\left(j\right),j=1,...,T_{p}$
to obtain an estimation $\hat{\mathbf{\Theta}}^{'}$of the covariance
matrix of $\mathbf{h}^{'}$.

\textbf{Step 4:} Output the estimation of the covariance matrix of
$\mathbf{h}$ as: $\hat{\mathbf{\Theta}}=\mathbf{U}\hat{\mathbf{\Theta}}^{'}\mathbf{U}^{\dagger}$.\medskip{}

Then the statistical CSI acquisition is summarized as follows. Within
each $T$ time slots, choose $T_{p}$ time slots such that the gap
between any two selected time slots is larger than channel coherent
time. Then each BS transmits $M$ dedicated pilot symbols in each
of the selected $T_{p}$ time slots for the users to estimate the
spatial correlation matrices using the rank deficient OAS estimator.
Finally, user $k$ sends the non-zero eigenvalues and the corresponding
eigenvectors of $\mathbf{\Theta}_{k,l},\forall l$ to the associated
BS.

\subsection{Real-time CSI Acquisition}

The dimension of the effective channel vector of a user in the $n$-th
user cluster is equal to $S_{n}$ ( the rank of the $n$-th Tx subspace
control variable $\mathbf{F}_{n}$), which is $O\left(1\right)$ and
is much less than $M$. To estimate the effective CSI $\widetilde{\mathbf{H}}_{n}$
of the $n$-th user cluster, the pilot symbols need to be precoded
using the $n$-th Tx subspace control variable $\mathbf{F}_{n}$ since
the effective CSI $\widetilde{\mathbf{H}}_{n}$ is restricted in the
subspace spanned by $\mathbf{F}_{n}$. As a result, we can simultaneously
transmit $N$ pilot symbols at a time for $N$ user clusters because
the Tx subspace control $\mathbf{F}$ can also mitigate the inter-cluster
interference in the pilot transmission stage. The total number of
orthogonal pilot symbols required for effective CSI estimation $\widetilde{\mathbf{H}}_{n},n=1,...,N$
is $\max_{n}S_{n}=O\left(1\right)$. Hence, by treating $\mathbf{F}_{n}$
as part of the effective channel, the estimation of the effective
CSI $\widetilde{\mathbf{H}}_{n}$ is similar to the conventional channel
estimation problem in a single-cell small-scale MIMO system and the
conventional CSI signaling method in LTE can be used to estimate the
effective CSI.

\subsection{Statistical and Real-time CSI Signaling Overhead}

The CSI signaling overhead includes the pilot symbol overhead (the
average number of orthogonal pilot symbols for real-time and statistical
CSI estimation) and the uplink CSI feedback overhead (the average
number of feedback vectors with different dimensions). For simplicity,
assume that each BS is associated with $K_{0}$ users, each user is
interfered by $L_{0}-1$ neighbor BSs, and $S_{n}=S,\forall n$. Then
the real-time CSI signaling overhead of the proposed scheme is ``$S$
PS, $K_{0}$ $\mathbb{C}^{S}$'' per time slot per cell, which means
that, the proposed scheme requires transmitting $S$ independent pilot
symbols and feed backing $K_{0}$ complex channel vectors of dimension
$S$ per time slot per cell. For each spatial correlation matrix,
the corresponding user only needs to send $R$ non-zero eigenvalues
and the corresponding eigenvectors to the BS. Hence, the statistical
CSI signaling overhead of the proposed scheme is ``$\frac{MT_{p}}{T}$
PS, $\frac{K_{0}L_{0}R}{T}$ $\mathbb{C}^{M}$, $\frac{K_{0}L_{0}}{T}$
$\mathbb{R}^{R}$'' per time slot per cell, which means that, the
proposed scheme requires transmitting $\frac{MT_{p}}{T}$ independent
pilot symbols, feed backing $\frac{K_{0}L_{0}R}{T}$ complex vectors
of dimension $M$ (eigenvectors of spatial correlation matrices),
and feed backing $\frac{K_{0}L_{0}}{T}$ real vectors of dimension
$R$ (eigenvalues of spatial correlation matrices) per time slot per
cell. As a comparison, the CSI signaling overhead of the conventional
single-cell MU-MIMO precoding scheme is ``$M$ PS, $K_{0}$ $\mathbb{C}^{M}$''
per time slot per cell. Since $T\gg T_{p}=O\left(R\right)=O\left(L_{0}\right)=O\left(1\right)$,
the overall CSI signaling overhead of the proposed scheme is significantly
smaller as will be shown in the simulations.

\subsection{Backhaul Signaling Overhead}

Suppose there are a total number of $N_{L}$ links (including both
direct links and cross links) in the network. The BSs need to send
$R$ non-zero eigenvalues and the corresponding eigenvectors of every
spatial correlation matrix to the central node within each $T$ time
slots. Hence, the backhaul signaling overhead is ``$\frac{N_{L}R}{T}$
$\mathbb{C}^{M}$, $\frac{N_{L}R}{T}$ $\mathbb{R}$'' per time slot
(i.e., $\frac{N_{L}R}{T}$ complex vectors of dimension $M$ and $\frac{N_{L}R}{T}$
real numbers per time slot). Since $T\gg R$ is usually true for massive
MIMO systems, the backhaul signaling overhead can be greatly reduced
compared to the cooperative MIMO or centralized coordinated MIMO where
the BSs need to send $N_{L}$ complex channel vectors of dimension
$M$ to the central node at each time slot.

\subsection{Summary of Practical Real-time Implementation Aspects}

In the proposed two stage precoding technique, there are four components
(steps) working at different timescales.

\textbf{Step 1 (Slow Statistical CSI Acquisition):} First, the BSs
need to obtain the spatial channel correlation matrices using the
method in Section \ref{sub:Slow-Statistical-CSI}, and then send them
to the central node. 

\textbf{Step 2 (Tx Subspace Control Optimization): }With the obtained
statistical CSI from the BSs, the central node calculates the optimal
Tx subspace control variables $\left\{ \mathbf{F}_{n}\right\} $ using
Algorithm BCA and sends $\mathbf{F}_{n}$ to BS $\overline{l}_{n}$
(which is the BS associated with the $n$-th user cluster). 

\textbf{Step 3 (Real-time CSI Acquisition):} By treating $\mathbf{F}_{n}$
as part of the effective channel, the estimation of the real-time
effective CSI $\widetilde{\mathbf{H}}_{n}$ is similar to the conventional
channel estimation problem in a single-cell small-scale MIMO system.

\textbf{Step 4 (Inner Precoding):} At each time slot, BS $\overline{l}_{n}$
performs inner precoding using the obtained effective CSI $\widetilde{\mathbf{H}}_{n}$.

Both the statistical CSI acquisition and the Tx subspace control optimization
can be done at a much slower timescale compared to the time slot duration.
Hence, the proposed solution is not sensitive to the backhaul signaling
latency delay and computational delay caused by the Tx subspace control
optimization. On the other hand, the effective CSI acquisition and
inner precoding per cluster is similar to that in the single-cell
small-scale MIMO system and thus they can be implemented in real-time
at each time slot. Note that in practice, it is impossible to obtain
perfect effective CSI at each BS due to the channel estimation/feedback
error. The impact of imperfect effective CSI on the performance is
similar to that in the conventional small-scale MIMO systems with
ZF precoding. As a result, we can also employ MMSE techniques for
the inner precoder to mitigate the effects of CSI errors. The theoretical
analysis of the effect of imperfect effective CSI on the two stage
subspace constrained precoding is left as an interesting future work.
\begin{rem}
[Extension to Frequency Selective Fading Channels]For clarity, we
assume flat fading channel in this paper. In the wideband systems
(such as OFDM) with frequency selective channels, the spatial correlation
matrices are approximately identical on different subcarriers \cite{Chandran_Springer04_AdaptiveAntenna,Sadek_TOC08_MIMO-OFDM}.
As a result, the same Tx subspace control $\mathbf{F}$ can be applied
to different subcarriers and the results and algorithm can also be
extended to the wideband systems. The extension to wideband systems
depends on the specific QoS requirements. If we require that the SINR
of each user on each subcarrier must be larger than a threshold with
high probability, then the proposed solution can be directly applied.
However, if we require that the rate of each user (rate summed over
all subcarriers) must be larger than a threshold with high probability,
then the extension may be non-trivial.
\end{rem}

\section{Simulation Results\label{sec:Simulation-Results}}

\subsection{Simulation Setup}

Consider a massive MIMO cellular system with $19$ cells, where a
reference cell is surrounded by two tiers of (interfering) cells.
The inter-site distance is $500$m. In each cell, there are $K_{0}$
users and $2$ uniformly distributed hotspots (user clusters). Each
hotspot contains $K_{c}$ users and the other $K_{0}-2K_{c}$ users
are uniformly distributed within the cell. Each BS is equipped with
$M$ antennas. The path gains $L_{k,l}$'s are first generated using
the path loss model (\textquotedblleft Urban Macro NLOS\textquotedblright{}
model) in \cite{3gpp_Rel9}. Then the spatial correlation matrices
$\mathbf{\Theta}_{k,l}$'s are generated according to the local scattering
model in \cite{Abdi_JSAC02_localscatering,Zhang_TWC07_localscattering}.
Specifically, the correlation between the $i,m$-th channel coefficients
of $\mathbf{h}_{k,l}$ is given by
\begin{equation}
\left[\mathbf{\Theta}_{k,l}\right]_{i,m}=\int_{\varphi_{k,l}^{\textrm{min}}}^{\varphi_{k,l}^{\textrm{max}}}\psi_{k,l}\left(\varphi\right)e^{j\varpi_{l}^{i,m}\left(\varphi\right)}d\varphi,\label{eq:SCmodel}
\end{equation}
where $\varphi$ denotes the AoD, $\psi_{k,l}\left(\varphi\right)$
denotes the power gain for the path specified by the AoD $\varphi$
and $\varpi_{l}^{i,m}\left(\varphi\right)$ is the phase difference
between the $i$-th and $m$-th BS antennas on the path specified
by $\varphi$. In the simulations, $\varpi_{l}^{i,m}\left(\varphi\right)$
in (\ref{eq:SCmodel}) is generated using a uniform linear array with
the antenna spacing equal to half wavelength, and $\psi_{k,l}\left(\varphi\right)=\overline{\psi}_{k,l},\forall\varphi\in\left[\varphi_{k,l}^{\textrm{min}},\varphi_{k,l}^{\textrm{max}}\right]$,
where $\overline{\psi}_{k,l}$ is chosen such that $\textrm{Tr}\left(\mathbf{\Theta}_{k,l}\right)=ML_{k,l}$
and $\varphi_{k,l}^{\textrm{min}},\varphi_{k,l}^{\textrm{max}}$ are
randomly generated such that 1) $\left|\varphi_{k,l}^{\textrm{max}}-\varphi_{k,l}^{\textrm{min}}\right|=\frac{\pi}{6}$
, 2) if user $k$ and user $k^{'}$ belong to the same hotspot, we
have $\varphi_{k,l}^{\textrm{min}}=\varphi_{k^{'},l}^{\textrm{min}},\varphi_{k,l}^{\textrm{max}}=\varphi_{k^{'},l}^{\textrm{max}}$.
We assume the spatial channel correlation matrices $\mathbf{\Theta}$
change every 1000 time slots.

In the \textit{basic simulation setup}, the per BS transmit power
is $P_{b}=10$dB and the other system parameters are set as $K_{0}=7$,
$K_{c}=3$, $M=40$, $\epsilon_{k}=0.05,\forall k$ and $w_{k}=L_{k,b_{k}},\forall k$;
the transmit powers of the users are given by $P_{k}=\frac{P_{b}}{K_{0}},\forall k$.
In the simulations, we will change the system parameters in the basic
setup to study the impact of different system parameters on the performance.
We only evaluate the performance of the reference cell to avoid ``unsymmetric
edge effects''. 

We compare the performance of the proposed solution with the following
baselines.
\begin{itemize}
\item \textbf{Baseline 1 (FFR)}: Fractional frequency reuse (FFR) \cite{Lei_PIMRC07_FFR}
is applied to suppress the inter-cell interference. There are 6 subbands,
3 of which are used for the inner zone and 3 are used for the outer
zone. The radius for the inner zone is set as 150m. In each cell,
ZF beamforming is used to serve the users on each subband. 
\item \textbf{Baseline 2 (Clustered CoMP)}: 3 neighbor BSs form a BS cluster
and employ cooperative ZF \cite{somekh2009cooperative} to simultaneously
serve all the users within the BS cluster.
\item \textbf{Baseline 3 (JSDM-PGP)}: This is the two-stage precoding scheme
proposed in \cite{Caire_TIT13_JSDM}, where the per-group processing
(PGP) with approximate BD is employed at each BS. In the approximate
BD, the pre-beamforming matrices $\mathbf{F}$ are chosen to satisfy
$\mathbf{F}_{n}^{\dagger}\mathbf{U}_{k,\overline{l}_{n}}^{\star}=0,\:\forall n\in\mathcal{B}_{k},\forall k$,
where $\mathbf{U}_{k,l}^{\star}\in\mathbb{U}^{M\times r_{k,l}^{\star}}$
is a semi-unitary matrix collecting the dominant eigenvectors of $\mathbf{\Theta}_{k,l}$.
The number $r_{k,l}^{\star}$ of dominate eigenvectors is chosen such
that $\lambda_{k,l}^{dB}\left(r_{k,l}^{\star}+1\right)\leq-20\textrm{dB}<\lambda_{k,l}^{dB}\left(r_{k,l}^{\star}\right)$,
where $\lambda_{k,l}^{dB}\left(m\right)$ is the $m$-th largest eigenvalue
of $\mathbf{\Theta}_{k,l}$ in dB. Simulations show that such choice
of $r_{k,l}^{\star}$ can achieve a good performance. Both ZF inner
precoder and regularized zero-forcing (RZF) \cite{Peel_TOC05_RCI}
inner precoder will be simulated.
\end{itemize}

\subsection{Verify the SINR Satisfaction Probability}

\begin{figure}
\begin{centering}
\includegraphics[width=85mm]{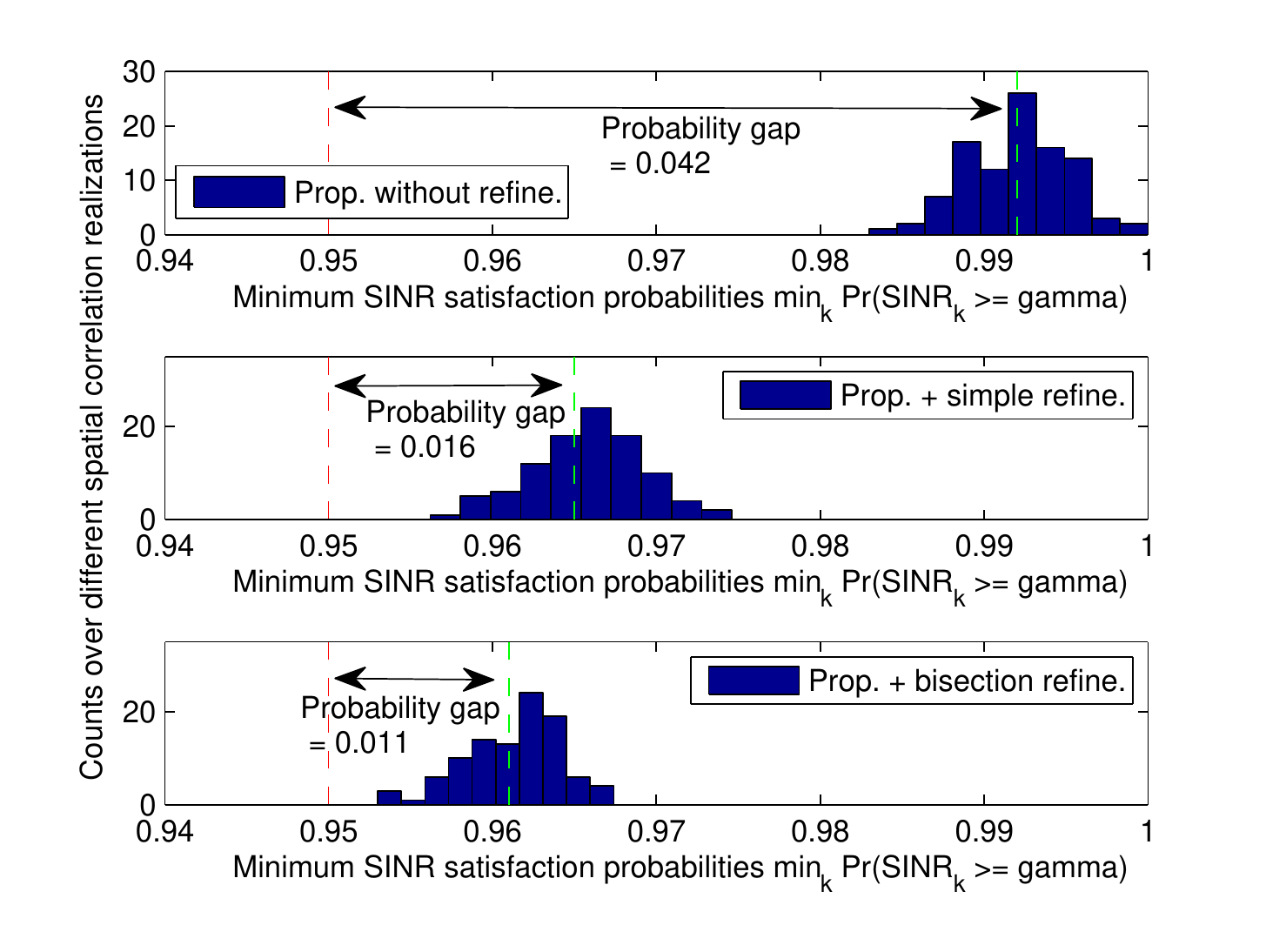}
\par\end{centering}

\protect\caption{\label{fig:SINR-Prob}{\small{}Histograms of the minimum SINR satisfaction
probabilities under the basic simulation setup. The red dash line
indicates the target SINR satisfaction probability and the green dash
line indicates the mean minimum SINR satisfaction probability.}}
\end{figure}

First, we verify that the original QoS requirement in (\ref{eq:ProbSINR})
is satisfied by the proposed solution under finite $M$. Note that
the QoS requirement in (\ref{eq:ProbSINR}) means that the \textit{minimum
SINR satisfaction probability} $\min_{k}\Pr\left\{ \textrm{SINR}_{k}\geq\gamma\right\} $
(conditioned on a given spatial correlation matrices $\mathbf{\Theta}$)
is larger than some target value $1-\epsilon_{k}$. Fig. \ref{fig:SINR-Prob}
shows the histograms of the minimum SINR satisfaction probabilities
over different realizations of $\mathbf{\Theta}$ under the basic
simulation setup ($M=40$). To obtain the histograms, we generated
100 realizations of $\mathbf{\Theta}$. Then, for each realization
of $\mathbf{\Theta}$, the minimum SINR satisfaction probability is
numerically evaluated using 10000 randomly generated realizations
of instantaneous CSI $\mathbf{H}$. Fig. \ref{fig:SINR-Prob} validates
that the proposed solution indeed achieve a minimum SINR satisfaction
probability no less than the required target $95\%$. The ``probability
gap'' in Fig. \ref{fig:SINR-Prob} refers to the difference between
the mean minimum SINR satisfaction probability and the target SINR
satisfaction probability. Without using the refinement methods, the
proposed solution is conservative, e.g., the ``probability gap''
is $0.042$. However, with the simple non-iterative refinement method,
the proposed solution become ``tighter'' (i.e., the ``probability
gap'' is smaller), and with the more complicated bisection refinement
method in \cite{Wang_ICASSP11_Bernstein-type}, the proposed solution
can be even tighter.

\subsection{Comparison of Outage Throughput Performance under Different System
Parameters}

\begin{figure}
\begin{centering}
\includegraphics[width=85mm]{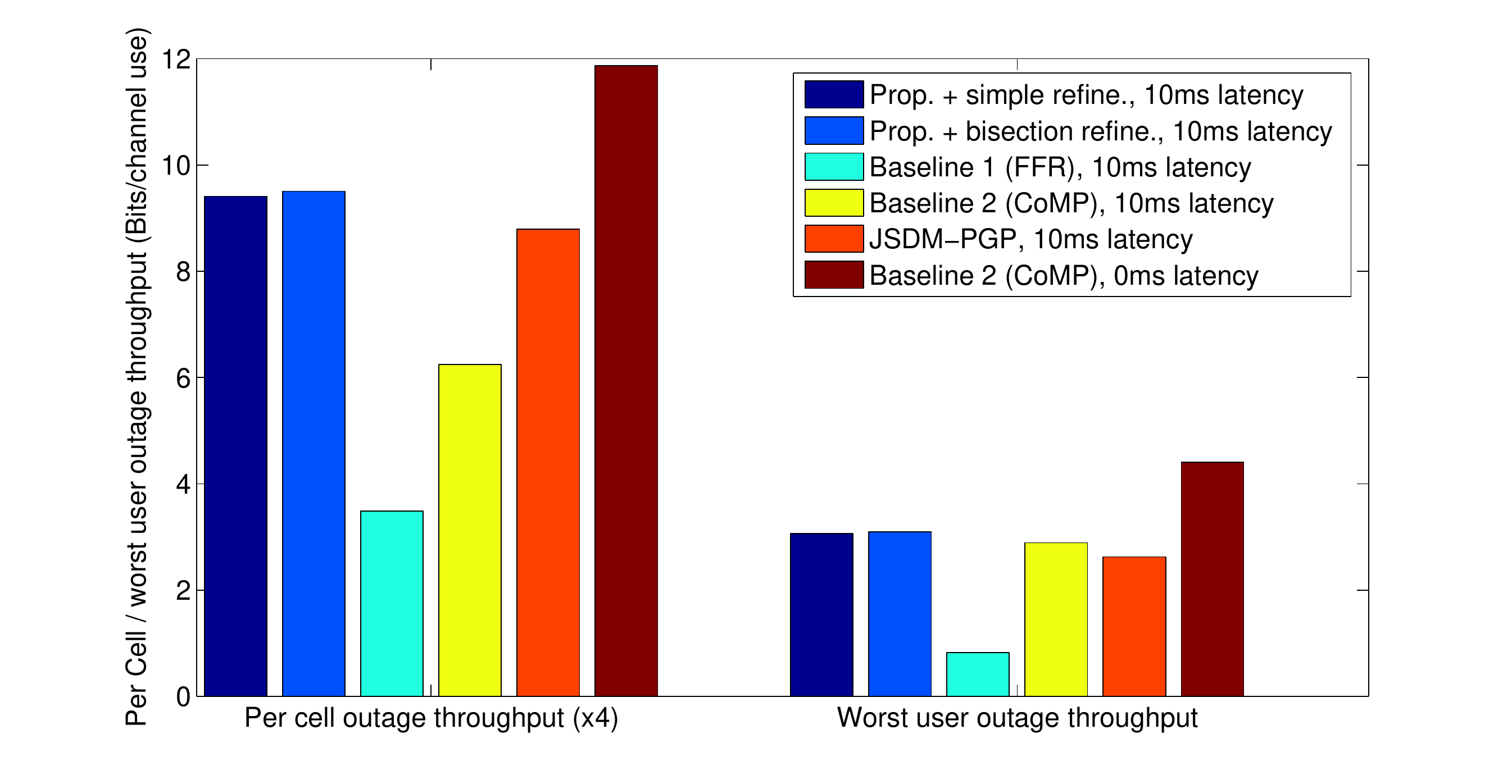}
\par\end{centering}

\protect\caption{\label{fig:bar_throughput}{\small{}Outage throughput comparisons
of different schemes under the basic simulation setup.}}
\end{figure}

In Fig. \ref{fig:bar_throughput}, we plot the outage throughput of
different schemes under the basic simulation setup, where the outage
throughput of user $k$ is defined as the maximum data rate that can
be achieved with a probability no less than $1-\epsilon_{k}$. For
baseline 2, the $3$ cooperative BSs need to exchange CSI and payload
data, and thus there is CSI delay when the backhaul latency is not
zero. When there is CSI delay, the outdated CSI is related to the
actual CSI by the autoregressive model in \cite{Baddour_TWC05_CSIdelaymodel}
with the following parameters: the user speed is 3 km/h; the carrier
frequency is $2$GHz. For baseline 3, ZF inner precoder is simulated.
It can be seen that the performance of the proposed scheme with the
simple non-iterative refinement method is close to that with the more
complicated bisection refinement method in \cite{Wang_ICASSP11_Bernstein-type}.
The outage throughput of the proposed scheme is larger than baseline
1 and baseline 3. Although the performance of baseline 2 is promising
at zero backhaul latency, the performance quickly degrades at 10ms
backhaul latency.

\begin{figure}
\begin{centering}
\includegraphics[width=85mm]{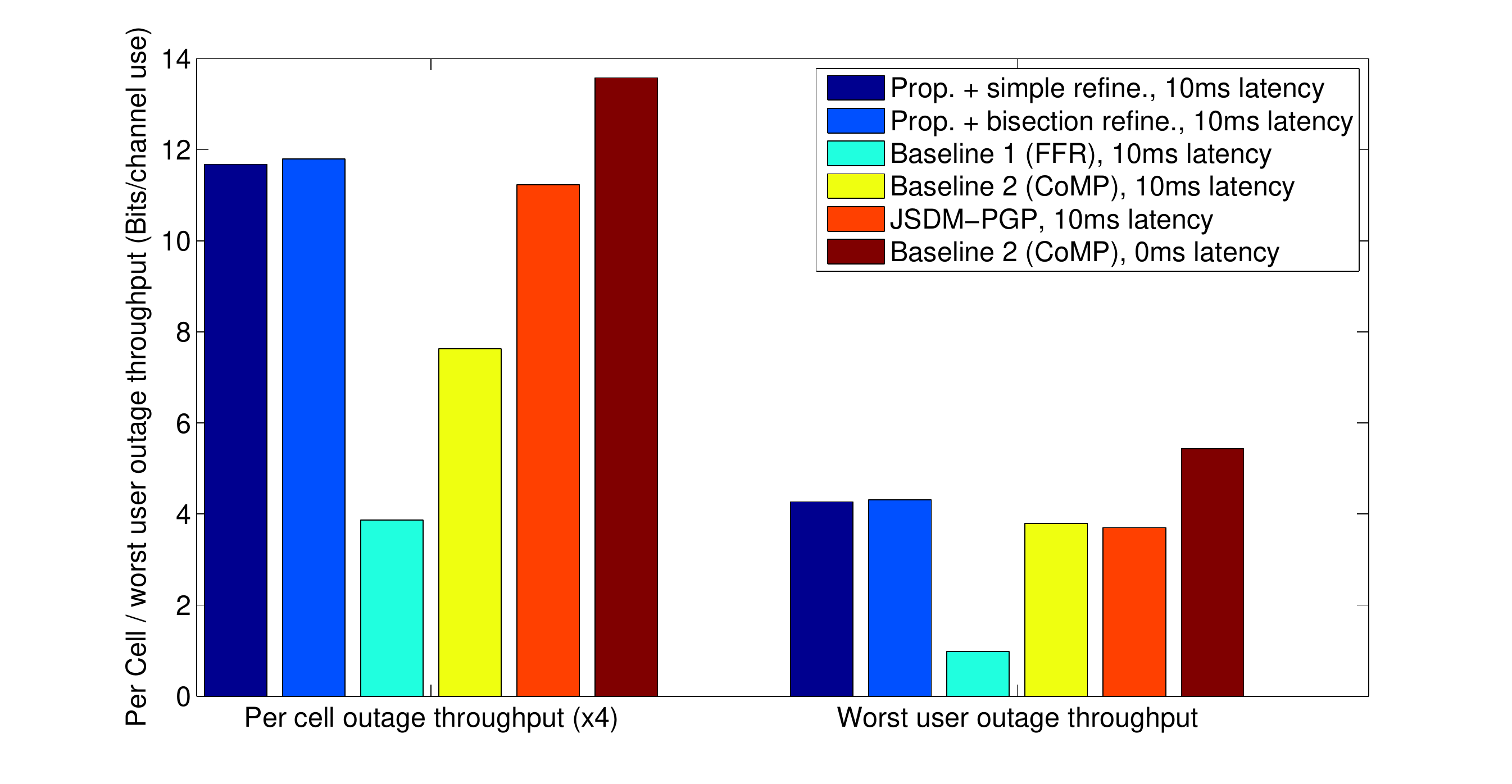}
\par\end{centering}

\protect\caption{\label{fig:bar_throughput2}{\small{}Outage throughput comparisons
of different schemes. The number of antennas is $M=64$ and the other
simulation setup is the same as the basic simulation setup.}}
\end{figure}

\begin{figure}
\begin{centering}
\includegraphics[width=85mm]{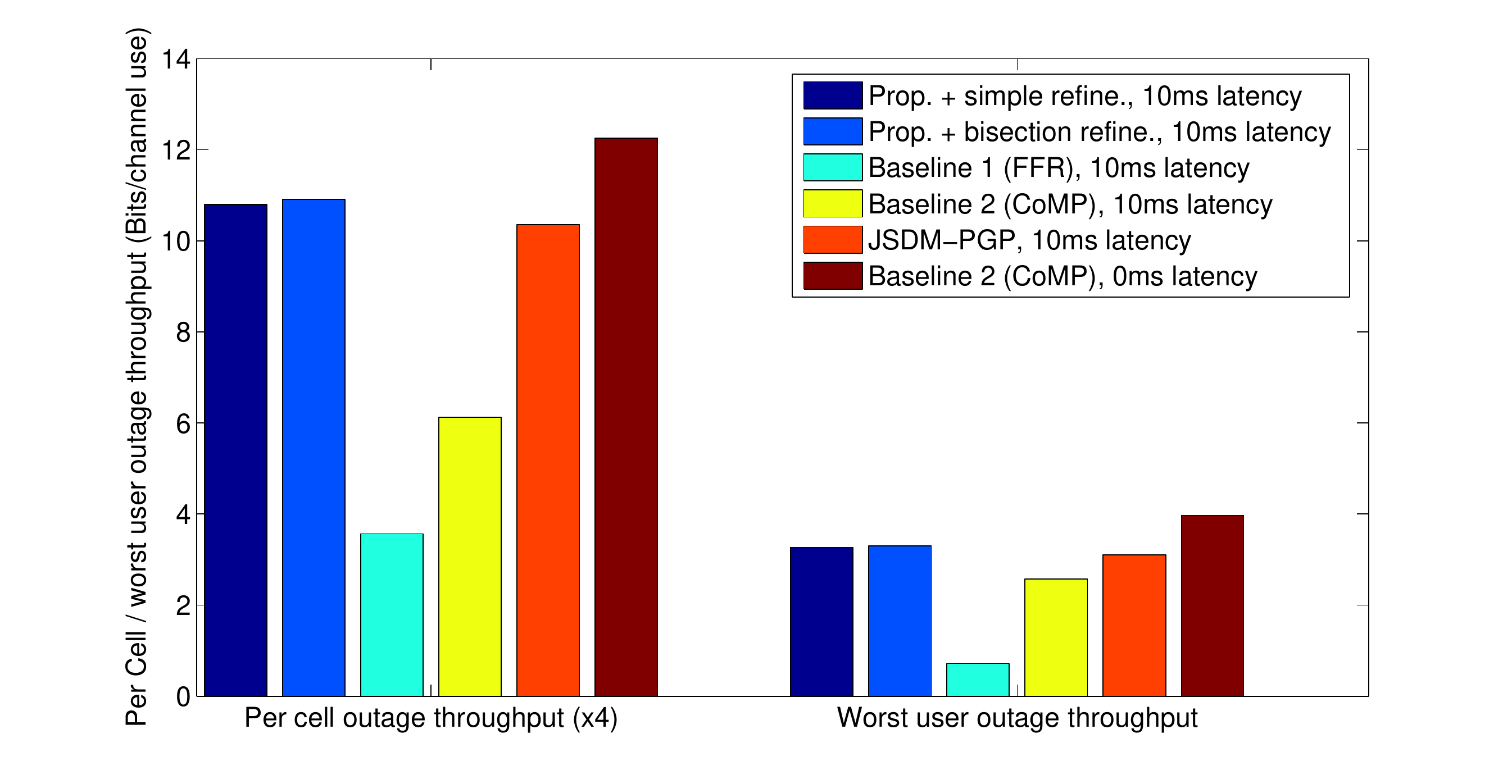}
\par\end{centering}

\protect\caption{\label{fig:bar_throughput3}{\small{}Outage throughput comparisons
of different schemes. The number of users in each cell is $K_{0}=8$
and the number of users per cluster is $K_{c}=4$. The other simulation
setup is the same as the basic simulation setup.}}
\end{figure}

\begin{figure}
\begin{centering}
\includegraphics[width=85mm]{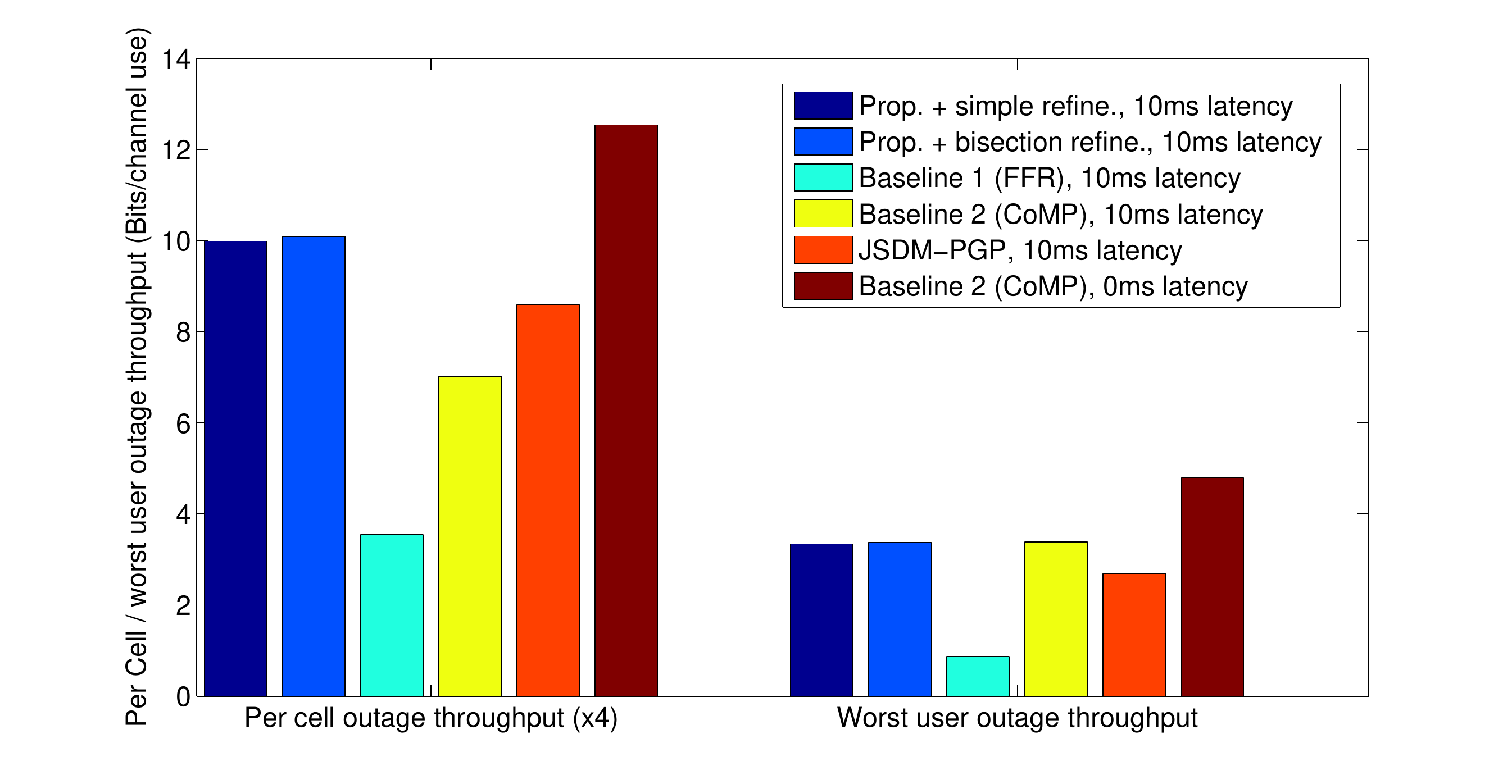}
\par\end{centering}

\protect\caption{\label{fig:bar_throughput4}{\small{}Outage throughput comparisons
of different schemes with heterogeneous QoS requirements. The other
simulation setup is the same as the basic simulation setup.}}
\end{figure}

In Fig. \ref{fig:bar_throughput2}, we change the number of antennas
to $M=64$. The other simulation setup is the same as the basic simulation
setup. Compared to Fig. \ref{fig:bar_throughput} with $M=40$ antennas,
the performance of all schemes improves as the number of antennas
increases. However, the performance improvement of the proposed scheme
and baseline 3 is larger. This is not surprising since the proposed
scheme and baseline 3 are specifically designed for massive MIMO systems
with spatially correlated channel. 

In Fig. \ref{fig:bar_throughput3}, we change the number of users
in each cell to $K_{0}=8$ and the number of users per cluster to
$K_{c}=4$. The other simulation setup is the same as the basic simulation
setup. In this case, the users in each cell concentrate in the two
hotspots. Compared to Fig. \ref{fig:bar_throughput}, the performance
of both the proposed scheme and baseline 3 are improved. This shows
that the proposed scheme and baseline 3 favor the scenario when the
users in each cell concentrate in a few user clusters. 

In Fig. \ref{fig:bar_throughput4}, we consider heterogeneous QoS
requirements, where $\epsilon_{k}$ is chosen uniformly between $\left[0.05,0.15\right]$.
The other simulation setup is the same as the basic simulation setup.
Compared to Fig. \ref{fig:bar_throughput} with homogeneous QoS requirement,
the performance gain of the proposed scheme over baseline 1 and 3
becomes larger. This shows that the proposed solution can better handle
the heterogeneous QoS requirements.

Under all simulation setups, the proposed scheme outperforms baseline
1 and baseline 3, as well as baseline 2 under the practical scenario
with 10ms backhaul latency. These results demonstrate the superior
performance and the robustness of the proposed optimization based
subspace constrained precoding w.r.t. signaling latency in backhaul.

\subsection{Impact of Inner Percoders and CSI Errors\label{sub:Impact-of-Inner}}

\begin{figure}
\begin{centering}
\includegraphics[width=85mm]{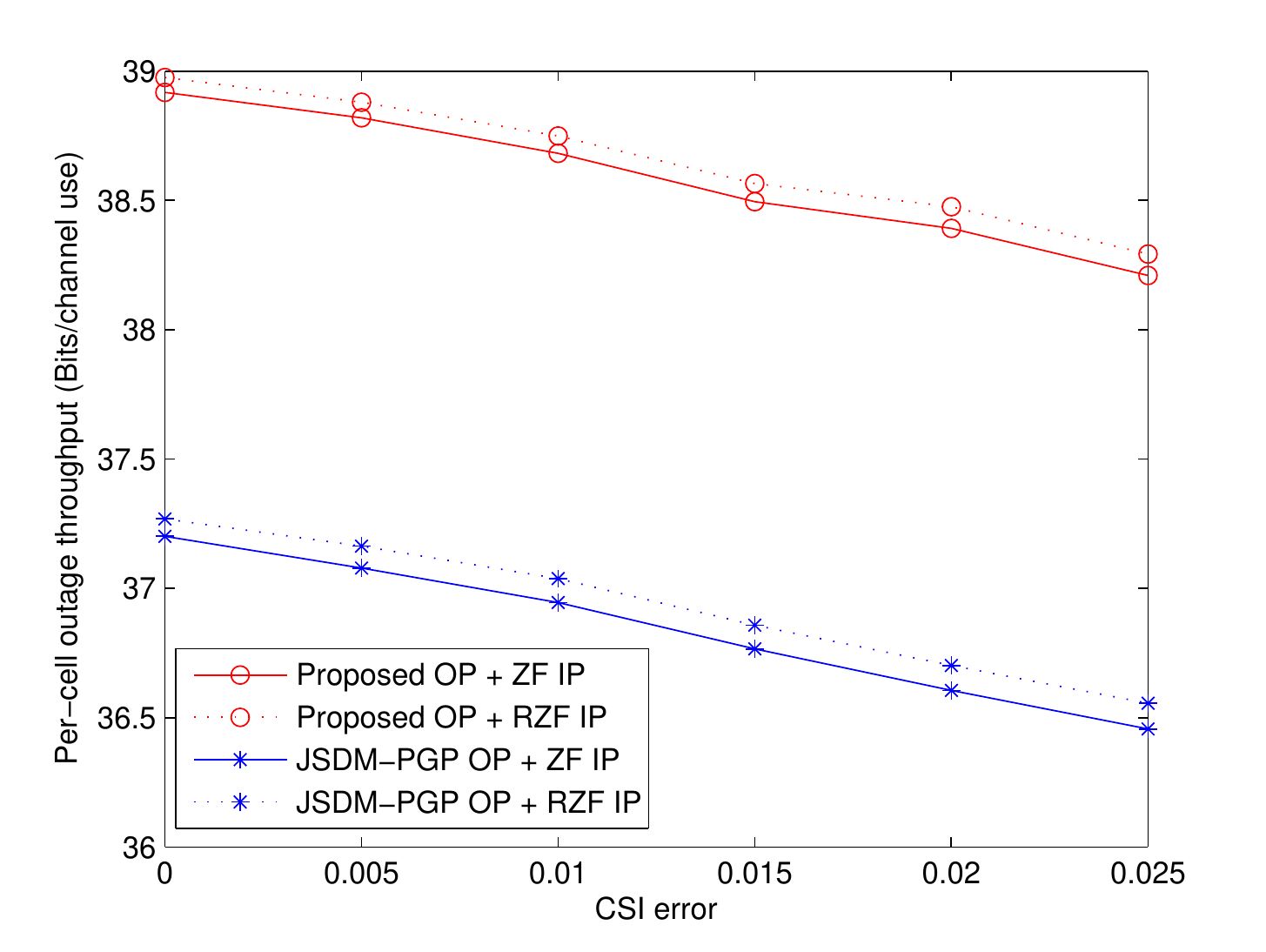}
\par\end{centering}

\protect\caption{\label{fig:bar_throughput-CSIerror}{\small{}An illustration of the
impact of inner precoding and CSI errors under the basic simulation
setup. The abbreviation ``OP'' stands for outer precoding and ``IP''
stands for inner precoding.}}
\end{figure}

In this paper, we consider ZF inner precoding because 1) the ZF inner
precoding is more tractable; and 2) it has been shown in \cite{Marzetta_SPM12_LargeMIMO}
that ZF precoding is close to optimal for massive MIMO systems with
perfect CSIT. To study the impact of different inner precoders, we
compare the performance of the ZF inner precoding with the RZF inner
precoding under different \textit{outer precoders} $\mathbf{F}$,
where the outer precoder refers to either the proposed optimization
based Tx subspace control matrix or the pre-beamforming matrix in
baseline 3. In RZF inner precoder, the regularization factor is fixed
to $\frac{K_{0}}{S_{0}P}$ as in \cite{Caire_TIT13_JSDM}, where $S_{0}$
is the total dimension of the outer precoders associated with the
user clusters in the reference cell. To study the impact of CSI errors,
the CSIT at each BS is modeled as follows: 
\[
\hat{\mathbf{h}}_{k,l}=\widetilde{\mathbf{h}}_{k,l}-\mathbf{e}_{k,l},
\]
where the CSI error $\mathbf{e}_{k,l}$ is assumed to be a complex
Gaussian vector with zero mean and covariance matrix $\sigma_{e}^{2}\mathbf{I}$,
and $\sigma_{e}^{2}$ denotes the error variance. In Fig. \ref{fig:bar_throughput-CSIerror},
we plot the outage throughput of different schemes versus the error
variance $\sigma_{e}^{2}$ under the basic simulation setup. It can
be seen that the performance of ZF inner precoding under the same
outer precoder is similar to that of RZF inner precoding. The performance
of all schemes degrades at a similar rate as the CSI error increases.
In all cases, the proposed optimization based outer precoder outperforms
the BD-based out percoder in baseline 3.

\subsection{Implementation Cost}

Table \ref{tab:Cputime} compares the implementation cost for different
schemes in terms of the computational complexity (CPU time), the (real-time
and statistical) CSI signaling overhead, and the backhaul signaling
overhead caused by the exchange of (real-time or statistical) CSI
between the central node and BSs. It can be seen that the overall
implementation cost of the proposed scheme and baseline 3 is the lowest
among all schemes. 

\begin{table}
\begin{centering}
{\footnotesize{}}%
\begin{tabular}{|l|l|l|l|l|}
\hline 
 & {\footnotesize{}CPU } & {\footnotesize{}Backhaul } & {\footnotesize{}Real-time } & {\footnotesize{}Statistical }\tabularnewline
 & {\footnotesize{}time} & {\footnotesize{}signaling} & {\footnotesize{}CSI } & {\footnotesize{}CSI}\tabularnewline
 &  &  & {\footnotesize{}signaling } & {\footnotesize{}signaling}\tabularnewline
\hline 
{\footnotesize{}Proposed} & {\footnotesize{}0.13 s} & {\footnotesize{}0.13 $\mathbb{C}^{40}$} & {\footnotesize{}9 PS, 7 $\mathbb{C}^{9}$} & {\footnotesize{}0.8 PS, 0.12 $\mathbb{C}^{40}$}\tabularnewline
{\footnotesize{}FFR} & {\footnotesize{}0.06 s} & {\footnotesize{}0} & {\footnotesize{}40 PS, 7 $\mathbb{C}^{40}$} & {\footnotesize{}0}\tabularnewline
{\footnotesize{}Ideal CoMP} & {\footnotesize{}1.29 s} & {\footnotesize{}16 $\mathbb{C}^{40}$} & {\footnotesize{}40 PS, 7 $\mathbb{C}^{120}$} & {\footnotesize{}0}\tabularnewline
{\footnotesize{}JSDM-PGP} & {\footnotesize{}0.04 s} & {\footnotesize{}0.13 $\mathbb{C}^{40}$} & {\footnotesize{}9 PS, 7 $\mathbb{C}^{9}$} & {\footnotesize{}0.8 PS, 0.12 $\mathbb{C}^{40}$}\tabularnewline
\hline 
\end{tabular}
\par\end{centering}{\footnotesize \par}

{\small{}\protect\caption{\label{tab:Cputime}{\small{}Comparison of the per time slot MATLAB
computational time and per time slot per cell signaling overhead of
different schemes under the basic simulation setup. For the proposed
scheme and Baseline 1, the statistical CSI $\mathbf{\Theta}$ is obtained
using the method described in Section \ref{sub:Slow-Statistical-CSI}
with $T_{p}=20$. The meaning of the CSI signaling overhead notation
in the table is also explained in Section \ref{sub:Slow-Statistical-CSI}.}}
}
\end{table}

\section{Conclusion\label{sec:Conlusion}}

We propose a two-stage subspace constrained precoding scheme for massive
MIMO cellular systems. The MIMO precoder is partitioned into \textit{inner
precoder} (for intra-cluster spatial multiplexing gain) and \textit{Tx
subspace control} matrix (for inter-cluster interference control).
We formulate the Tx subspace control as a Quality of Service (QoS)
optimization problem and propose a novel bi-convex approximation approach
to produce efficiently computable bi-convex approximation of this
QoS optimization problem. Analysis shows that the proposed solution
is robust to backhaul signaling latency and can significantly reduce
the CSI signaling overhead. Moreover, simulations verify that the
proposed design achieves significant performance gain compared with
various state-of-the-art baselines under various backhaul signaling
latency.

\appendix

\subsection{Proof of Theorem \ref{thm:Iapprox}\label{sub:Proof-of-TheoremIapprox}}

Note that $I_{k}=\sum_{n\in\mathcal{B}_{k}}M\overline{\mathbf{z}}_{k,\overline{l}_{n}}^{\dagger}\mathbf{\Theta}_{k,\overline{l}_{n}}^{1/2}\mathbf{F}_{n}\mathbf{G}_{n}\mathbf{P}_{n}\mathbf{G}_{n}^{\dagger}\mathbf{F}_{n}^{\dagger}\mathbf{\Theta}_{k,\overline{l}_{n}}^{1/2}\overline{\mathbf{z}}_{k,\overline{l}_{n}}$,
where $\overline{\mathbf{z}}_{k,\overline{l}_{n}}\triangleq\frac{1}{\sqrt{M}}\mathbf{z}_{k,\overline{l}_{n}}\in\mathbb{C}^{M}$
has i.i.d. complex entries of zero mean and variance $\frac{1}{M}$.
It can be verified that the inner ZF precoder $\mathbf{G}_{n}$ satisfies
$\mathbf{G}_{n}\mathbf{G}_{n}^{\dagger}=\widetilde{\mathbf{H}}_{n}^{\dagger}\left(\widetilde{\mathbf{H}}_{n}\widetilde{\mathbf{H}}_{n}^{\dagger}\right)^{-1}\textrm{diag}\left(\left(\widetilde{\mathbf{H}}_{n}\widetilde{\mathbf{H}}_{n}^{\dagger}\right)^{-1}\right)^{-1}\left(\widetilde{\mathbf{H}}_{n}\widetilde{\mathbf{H}}_{n}^{\dagger}\right)^{-1}\widetilde{\mathbf{H}}_{n}$,
from which it follows that 
\[
\left\Vert \mathbf{G}_{n}\mathbf{P}_{n}\mathbf{G}_{n}^{\dagger}\right\Vert \leq M\overline{P}_{n}\widetilde{\lambda}_{n,\textrm{max}}\left\Vert \widetilde{\mathbf{H}}_{n}^{\dagger}\left(\widetilde{\mathbf{H}}_{n}\widetilde{\mathbf{H}}_{n}^{\dagger}\right)^{-2}\widetilde{\mathbf{H}}_{n}\right\Vert \leq\frac{\overline{P}_{n}\widetilde{\lambda}_{n,\textrm{max}}}{\widetilde{\lambda}_{n,\textrm{min}}},
\]
where $\widetilde{\lambda}_{n,\textrm{max}}$ and $\widetilde{\lambda}_{n,\textrm{min}}$
are respectively the largest and the smallest eigenvalue of $\frac{1}{M}\widetilde{\mathbf{H}}_{n}\widetilde{\mathbf{H}}_{n}^{\dagger}$.
It follows from Assumption \ref{asm:LSFM} that $\underset{M\rightarrow\infty}{\limsup}\left\Vert \frac{1}{M}\mathbf{H}_{n}\mathbf{H}_{n}^{\dagger}\right\Vert <\infty$
with probability 1 \cite{Wagner_TIT12s_LargeMIMO}, where $\mathbf{H}_{n}=\left[\mathbf{h}_{k,\overline{l}_{n}}\right]_{k\in\mathcal{U}_{n}}^{\dagger}\in\mathbb{C}^{\left|\mathcal{U}_{n}\right|\times M}$.
Since $\widetilde{\lambda}_{n,\textrm{max}}=\left\Vert \frac{1}{M}\widetilde{\mathbf{H}}_{n}\widetilde{\mathbf{H}}_{n}^{\dagger}\right\Vert \leq\left\Vert \frac{1}{M}\mathbf{H}_{n}\mathbf{H}_{n}^{\dagger}\right\Vert $,
$\widetilde{\lambda}_{n,\textrm{max}}$ is uniformly bounded with
probability 1. On the other hand, there exists $\widetilde{\lambda}_{0}>0$
such that, we have $\widetilde{\lambda}_{n,\textrm{min}}\geq\widetilde{\lambda}_{0}$
uniformly on $M$; otherwise, it will contradict with the assumption
that $\mathbf{F}$ is feasible. It follows from the above analysis
that $\left\Vert \mathbf{G}_{n}\mathbf{P}_{n}\mathbf{G}_{n}^{\dagger}\right\Vert $
is uniformly bounded on $M$ with probability 1. Together with Assumption
\ref{asm:LSFM}, $\left\Vert \mathbf{\Theta}_{k,\overline{l}_{n}}^{1/2}\mathbf{F}_{n}\mathbf{G}_{n}\mathbf{P}_{n}\mathbf{G}_{n}^{\dagger}\mathbf{F}_{n}^{\dagger}\mathbf{\Theta}_{k,\overline{l}_{n}}^{1/2}\right\Vert $
is also uniformly bounded on $M$ with probability 1. Then by \cite[Lemma 4]{Wagner_TIT12s_LargeMIMO},
we have $I_{k}-\overline{I}_{k}\overset{a.s}{\rightarrow}0$, where
\begin{eqnarray*}
\overline{I}_{k} & = & \sum_{n\in\mathcal{B}_{k}}\textrm{Tr}\left(\mathbf{\Theta}_{k,\overline{l}_{n}}^{1/2}\mathbf{F}_{n}\mathbf{G}_{n}\mathbf{P}_{n}\mathbf{G}_{n}^{\dagger}\mathbf{F}_{n}^{\dagger}\mathbf{\Theta}_{k,\overline{l}_{n}}^{1/2}\right)\\
 & \leq & \sum_{n\in\mathcal{B}_{k}}\textrm{Tr}\left(\mathbf{G}_{n}\mathbf{P}_{n}\mathbf{G}_{n}^{\dagger}\right)\textrm{Tr}\left(\mathbf{F}_{n}^{\dagger}\mathbf{\Theta}_{k,\overline{l}_{n}}\mathbf{F}_{n}\right)=\hat{I}_{k},
\end{eqnarray*}
from which Theorem \ref{thm:Iapprox} follows immediately.

\subsection{Proof of Theorem \ref{thm:Conservative-Approximation} \label{sub:Proof-of-TheoremCA}}

Let $\widetilde{\mathbf{H}}_{-k}=\left[\widetilde{\mathbf{h}}_{k^{'},\overline{n}_{k}}\right]_{k^{'}\in\mathcal{U}_{\overline{n}_{k}}\backslash\left\{ k\right\} }^{\dagger}\in\mathbb{C}^{\left(\left|\mathcal{U}_{\overline{n}_{k}}\right|-1\right)\times S_{\overline{n}_{k}}}$
and consider the singular value decomposition (SVD): $\widetilde{\mathbf{H}}_{-k}=\widetilde{\mathbf{U}}\widetilde{\mathbf{D}}\mathbf{\widetilde{V}}^{\dagger}$,
where $\widetilde{\mathbf{U}}\in\mathbb{U}^{\left(\left|\mathcal{U}_{\overline{n}_{k}}\right|-1\right)\times\left(\left|\mathcal{U}_{\overline{n}_{k}}\right|-1\right)}$,
$\widetilde{\mathbf{D}}\in\mathbb{R}^{\left(\left|\mathcal{U}_{\overline{n}_{k}}\right|-1\right)\times\left(\left|\mathcal{U}_{\overline{n}_{k}}\right|-1\right)}$,
$\widetilde{\mathbf{V}}\in\mathbb{U}^{S_{\overline{n}_{k}}\times\left(\left|\mathcal{U}_{\overline{n}_{k}}\right|-1\right)}$.
Let $\widetilde{\mathbf{V}}_{c}\in\mathbb{U}^{S_{\overline{n}_{k}}\times\left(S_{\overline{n}_{k}}-\left|\mathcal{U}_{\overline{n}_{k}}\right|+1\right)}$
be the orthogonal complement of $\widetilde{\mathbf{V}}$, i.e., $\widetilde{\mathbf{V}}_{c}^{\dagger}\widetilde{\mathbf{V}}=\mathbf{0}$.
Using similar proof as that of \cite{Li_TIT06_MMSEsinrdist}, it can
be shown that $\left|\mathbf{h}_{k,b_{k}}^{\dagger}\mathbf{F}_{\overline{n}_{k}}\mathbf{g}_{k}\right|^{2}=\mathbf{h}_{k,b_{k}}^{\dagger}\mathbf{F}_{\overline{n}_{k}}\widetilde{\mathbf{V}}_{c}\widetilde{\mathbf{V}}_{c}^{\dagger}\mathbf{F}_{\overline{n}_{k}}^{\dagger}\mathbf{h}_{k,b_{k}}$.
Hence, conditioned on $\widetilde{\mathbf{H}}_{-k}$, $\widetilde{\mathbf{V}}_{c}$
is fixed, and thus $\mathbf{e}\triangleq\widetilde{\mathbf{V}}_{c}^{\dagger}\mathbf{F}_{\overline{n}_{k}}^{\dagger}\mathbf{h}_{k,b_{k}}\sim\mathcal{CN}\left(\mathbf{0},\mathbf{Q}\right)$
is a complex Gaussian vector with covariance $\mathbf{Q}=\widetilde{\mathbf{V}}_{c}^{\dagger}\mathbf{F}_{\overline{n}_{k}}^{\dagger}\mathbf{\Theta}_{\overline{n}_{k}}^{\circ}\mathbf{F}_{\overline{n}_{k}}\widetilde{\mathbf{V}}_{c}$.
Then conditioned on $\widetilde{\mathbf{H}}_{-k}$, $\left|\mathbf{h}_{k,b_{k}}^{\dagger}\mathbf{F}_{\overline{n}_{k}}\mathbf{g}_{k}\right|^{2}=\overline{\mathbf{e}}^{\dagger}\mathbf{Q}\overline{\mathbf{e}}$,
where $\overline{\mathbf{e}}$ is a standard complex Gaussian vector.
Applying the Bernstein-type inequality based method in \cite{Bechar_arxiv11_Bernstein-type,Wang_ICASSP11_Bernstein-type},
we have {\small{}
\begin{equation}
\Pr\left\{ \left.\left|\mathbf{h}_{k,b_{k}}^{\dagger}\mathbf{F}_{\overline{n}_{k}}\mathbf{g}_{k}\right|^{2}\geq\frac{w_{k}\gamma}{P_{k}}\left(\hat{I}_{k}+1\right)\right|\widetilde{\mathbf{H}}_{-k}\right\} \geq1-\epsilon_{k},\label{eq:BC0}
\end{equation}
}if the following sufficient condition is satisfied 
\begin{equation}
\textrm{Tr}\left(\overline{\mathbf{Q}}\right)-\sqrt{2\delta_{k}}\sqrt{\left\Vert \overline{\mathbf{Q}}\right\Vert _{F}^{2}}\geq\frac{w_{k}\gamma}{\lambda_{\overline{n}_{k}}P_{k}}\left(\hat{I}_{k}+1\right),\label{eq:BC1}
\end{equation}
where $\overline{\mathbf{Q}}=\frac{1}{\lambda_{\overline{n}_{k}}}\mathbf{Q}$.
It is easy to see that all the eigenvalues of $\overline{\mathbf{Q}}$
must be no more than 1. As a result, we have $\left\Vert \overline{\mathbf{Q}}\right\Vert _{F}^{2}\leq\textrm{Tr}\left(\overline{\mathbf{Q}}\right)$
and thus (\ref{eq:BC1}) can be satisfied if
\begin{equation}
\textrm{Tr}\left(\overline{\mathbf{Q}}\right)-\sqrt{2\delta_{k}}\sqrt{\textrm{Tr}\left(\overline{\mathbf{Q}}\right)}\geq\frac{w_{k}\gamma}{\lambda_{\overline{n}_{k}}P_{k}}\left(\hat{I}_{k}+1\right).\label{eq:BC2}
\end{equation}
It is easy to verify that (\ref{eq:BC2}) can be satisfied if
\begin{eqnarray}
\left(1-\frac{\sqrt{2\delta_{k}}}{\sigma_{k}}\right)\textrm{Tr}\left(\overline{\mathbf{Q}}\right) & \geq & \frac{w_{k}\gamma}{\lambda_{\overline{n}_{k}}P_{k}}\left(\hat{I}_{k}+1\right),\nonumber \\
\textrm{Tr}\left(\overline{\mathbf{Q}}\right) & \geq & \sigma_{k}^{2}.\label{eq:BC3}
\end{eqnarray}
It can be shown that $\textrm{Tr}\left(\overline{\mathbf{Q}}\right)\geq\sum_{j=\left|\mathcal{U}_{\overline{n}_{k}}\right|}^{S_{\overline{n}_{k}}}\overline{\lambda}_{j}$,
where $\overline{\lambda}_{j}$ is the $j$-th largest eigenvalue
of $\mathbf{F}_{\overline{n}_{k}}^{\dagger}\overline{\mathbf{\Theta}}_{\overline{n}_{k}}^{\circ}\mathbf{F}_{b_{k}}$.
Using the fact that $\overline{\lambda}_{j}\leq1,\forall j$, we have
$\sum_{j=\left|\mathcal{U}_{\overline{n}_{k}}\right|}^{S_{\overline{n}_{k}}}\overline{\lambda}_{j}\geq\textrm{Tr}\left(\overline{\mathbf{\Theta}}_{\overline{n}_{k}}^{\circ}\mathbf{F}_{\overline{n}_{k}}\mathbf{F}_{\overline{n}_{k}}^{\dagger}\right)-\left|\mathcal{U}_{\overline{n}_{k}}\right|+1$.
Hence $\textrm{Tr}\left(\overline{\mathbf{Q}}\right)\geq\textrm{Tr}\left(\overline{\mathbf{\Theta}}_{\overline{n}_{k}}^{\circ}\mathbf{F}_{\overline{n}_{k}}\mathbf{F}_{\overline{n}_{k}}^{\dagger}\right)-\left|\mathcal{U}_{b_{k}}\right|+1$.
Note that (\ref{eq:CC12}) is obtained from (\ref{eq:BC3}) by replacing
$\textrm{Tr}\left(\overline{\mathbf{Q}}\right)$ with $\textrm{Tr}\left(\overline{\mathbf{\Theta}}_{b_{k}}^{\circ}\mathbf{F}_{b_{k}}\mathbf{F}_{b_{k}}^{\dagger}\right)-\left|\mathcal{U}_{b_{k}}\right|+1$.
Hence, (\ref{eq:CC12}) ensures that (\ref{eq:BC3}) is satisfied.
Clearly, the above relationship between (\ref{eq:BC0}-\ref{eq:BC3})
and (\ref{eq:CC12}) holds for any $\widetilde{\mathbf{H}}_{-k}$.
As a result, (\ref{eq:CC12}) also ensures that the unconditional
version of (\ref{eq:BC0}) in (\ref{eq:CCapprox}) is satisfied.

\subsection{Proof of Theorem \ref{thm:Tightness-of-SDR} \label{Proof-of-TheoremSDR}}

It is easy to see that $\mathbf{W}^{\star}$ must also be the optimal
solution of $\hat{\mathcal{P}}$ with fixed $\gamma^{\star}$, which
is a convex problem w.r.t. $\mathbf{W}^{\star}$. According to the
analysis in Section \ref{sub:Lagrange-dual-method}, there exist Lagrange
multipliers $\boldsymbol{\mu}^{\star}$ and $\boldsymbol{\nu}^{\star}$
such that $\mathbf{W}^{\star}$ maximizes the corresponding Lagrange
function $L\left(\boldsymbol{\mu}^{\star},\boldsymbol{\nu}^{\star},\mathbf{W}\right)$
in (\ref{eq:DualFun}). The optimal solution $\mathbf{W}^{\star}$
of the maximization problem in (\ref{eq:DualFun}) with $\left(\boldsymbol{\mu},\boldsymbol{\nu}\right)=\left(\boldsymbol{\mu}^{\star},\boldsymbol{\nu}^{\star}\right)$
is given by (\ref{eq:optWn}) with $\left(\boldsymbol{\mu},\boldsymbol{\nu}\right)$
in $\mathbf{A}_{n}$ equal to $\left(\boldsymbol{\mu}^{\star},\boldsymbol{\nu}^{\star}\right)$.
From (\ref{eq:optWn}), it is easy to see that $\mathbf{W}_{n}^{\star}-\mathbf{W}_{n}^{\star2}=\mathbf{0},\forall n$,
from which Theorem \ref{thm:Tightness-of-SDR} follows immediately.%

\end{document}